\documentclass[twoside,twocolumn,9pt]{article}
\usepackage{extsizes}
\usepackage[super,sort&compress,comma]{natbib}
\usepackage[version=3]{mhchem}
\usepackage[left=1.5cm, right=1.5cm, top=1.785cm, bottom=2.0cm]{geometry}
\usepackage{balance}
\usepackage{times,mathptmx}
\usepackage{sectsty}
\usepackage{graphicx}
\usepackage{lastpage}
\usepackage[format=plain,justification=justified,singlelinecheck=false,font={stretch=1.125,small,sf},labelfont=bf,labelsep=space]{caption}
\usepackage{float}
\usepackage{fancyhdr}
\usepackage{fnpos}
\usepackage[british]{babel}
\usepackage{multirow} 
\usepackage{array}
\usepackage{droidsans}
\usepackage{charter}
\usepackage[T1]{fontenc}
\usepackage[usenames,dvipsnames]{xcolor}
\usepackage{setspace}
\usepackage[compact]{titlesec}
\newcommand{\tb}{\textsubscript}	
\newcommand{\matlab}{\textsc{Matlab}}	
\newcommand{\comsol}{\textsc{Comsol}}	
\usepackage{mathtools}	
\usepackage{amssymb}		
\usepackage{siunitx}			

\let\Re\relax								
\DeclareMathOperator\Re{Re}			
\let\temp\phi \let\phi\varphi \let\varphi\temp 						
\let\temp\epsilon \let\epsilon\varepsilon \let\varepsilon\temp	
\newcommand{\abs}[1]{\left|#1\right|}		
\newcommand{\mathsc}[1]{{\text{\normalfont\scshape{#1}}}}	

\newcommand{\sig}[2]{\ensuremath{\sigma_\mathrm{#1}^\mathsc{#2}}}

\newcommand{\vF}{\ensuremath{v_\textsc{f}}}

\usepackage{graphicx}		
	\graphicspath{{./Figs/}}	
\usepackage{acro}			
\DeclareAcronym{aptes}{short = APTES, long = 3-aminopropyl triethoxysilane}
\DeclareAcronym{bf}{short = BF, long = bright-field}
\DeclareAcronym{bfp}{short = BFP, long = back focal plane}
\DeclareAcronym{ccd}{short = CCD, long = charge-coupled device}
\DeclareAcronym{df}{short = DF, long = dark-field}
\DeclareAcronym{led}{short = LED, long = light-emitting diode}
\DeclareAcronym{lspr}{short = LSPR, long = localized surface plasmon resonance}
\DeclareAcronym{na}{short = NA, long = numerical aperture}
\DeclareAcronym{np}{short = NP, long = nanoparticle}
\DeclareAcronym{pw}{short = PW, long = plane-wave}
\DeclareAcronym{si}{short = SI, long = supporting information}
\DeclareAcronym{tem}{short = TEM, long = transmission electron microscopy}
\DeclareAcronym{dda}{short = DDA , long = discrete dipole approximation}
\usepackage{xr} 
\usepackage{xpatch} 
\makeatletter
\xpatchcmd{\@ssect@ltx}{\@xsect}{\protected@edef\@currentlabelname{#8}\@xsect}{}{}
\xpatchcmd{\@sect@ltx}{\@xsect}{\protected@edef\@currentlabelname{#8}\@xsect}{}{}
\makeatother

\usepackage{hyperref} 
\hypersetup{%
	linktocpage=true, colorlinks=true, allcolors= blue,%
	pdfstartpage=1, pdfstartview=FitV,%
	breaklinks=true, pdfpagemode=UseNone, pageanchor=true, pdfpagemode=UseOutlines,%
	plainpages=false, bookmarksnumbered, bookmarksopen=true, bookmarksopenlevel=2,%
	hypertexnames=true, pdfhighlight=/O,
	pdftitle={Quantitative Optical Cross Section Microspectroscopy of Individual Silver Nanocubes Reveals Surface Compositional Changes at the Nanoscale},%
	pdfauthor={Yisu Wang, Attilio Zilli, Zoltan Sztranyovszky, Wiebke Albrecht, Sara Bals, Paola Borri and Wolfgang Langbein},%
	pdfsubject={},%
	pdfkeywords={Nanoplasmonics, Nanoparticles, Optical cross sections, Optical–electron microscopy, Scattering microspectroscopy, Microscopy models},%
	pdfcreator={pdfLaTeX},%
	pdfproducer={LaTeX with hyperref package}%
	}%
\definecolor{cream}{RGB}{222,217,201}
\newcommand{\Onlinecite}[1]{ref.\nocite{#1}\citenum{#1}}
\addto\extrasenglish{  }

\newif\ifcomment
\usepackage{soul}
\ifcomment
\newcommand{\add}[1]{\textcolor{red}{#1}}
\newcommand{\delete}[1]{\textcolor{red}{\st{#1}}}

\newcommand{\mnote}[1]{\marginpar{\textcolor{green}{\textbf{#1}}}}
\else
\newcommand{\add}[1]{#1}
\newcommand{\delete}[1]{}

\newcommand{\mnote}[1]{}
\fi

\begin{document}
	\pagestyle{fancy} \thispagestyle{plain} \fancypagestyle{plain}{
		
		\renewcommand{\headrulewidth}{0pt}
	}%
%
	\makeFNbottom \makeatletter
	\renewcommand\LARGE{\@setfontsize\LARGE{15pt}{17}}
	\renewcommand\Large{\@setfontsize\Large{12pt}{14}}
	\renewcommand\large{\@setfontsize\large{10pt}{12}}
	\renewcommand\footnotesize{\@setfontsize\footnotesize{7pt}{10}}
	\makeatother
	\renewcommand{\thefootnote}{\fnsymbol{footnote}}
	\renewcommand\footnoterule{\vspace*{1pt}%
		\color{cream}\hrule width 3.5in height 0.4pt
		\color{black}\vspace*{5pt}} \setcounter{secnumdepth}{5}
	\makeatletter
	\renewcommand\@biblabel[1]{#1}
	\renewcommand\@makefntext[1]%
	{\noindent\makebox[0pt][r]{\@thefnmark\,}#1} \makeatother
	\renewcommand{\figurename}{\small{Fig.}~}
	\sectionfont{\sffamily\Large}
	\subsectionfont{\normalsize}
	\subsubsectionfont{\bf}
	\setstretch{1.125} 
	\setlength{\skip\footins}{0.8cm} \setlength{\footnotesep}{0.25cm}
	\setlength{\jot}{10pt}
	\titlespacing*{\section}{0pt}{4pt}{4pt}
	\titlespacing*{\subsection}{0pt}{15pt}{1pt}
%
	\fancyfoot[RO]{\footnotesize{\sffamily{1--\pageref{LastPage}
				~\textbar  \hspace{2pt}\thepage}}}
	\fancyfoot[LE]{\footnotesize{\sffamily{\thepage~\textbar\hspace{3.45cm}
				1--\pageref{LastPage}}}} \fancyhead{}
	\renewcommand{\headrulewidth}{0pt}
	\renewcommand{\footrulewidth}{0pt}
	\setlength{\arrayrulewidth}{1pt} \setlength{\columnsep}{6.5mm}
	\setlength\bibsep{1pt}
%
\makeatletter
\newlength{\figrulesep}
\setlength{\figrulesep}{0.5\textfloatsep}
\newcommand{\topfigrule}{\vspace*{-1pt}%
	\noindent{\color{cream}\rule[-\figrulesep]{\columnwidth}{1.5pt}} }
\newcommand{\botfigrule}{\vspace*{-2pt}%
	\noindent{\color{cream}\rule[\figrulesep]{\columnwidth}{1.5pt}} }
\newcommand{\dblfigrule}{\vspace*{-1pt}%
	\noindent{\color{cream}\rule[-\figrulesep]{\textwidth}{1.5pt}} }
\makeatother
%
\twocolumn[
\begin{@twocolumnfalse}
	\sffamily
	\begin{tabular}{p{18cm} }
		
		\noindent\LARGE{\textbf{Quantitatively linking morphology and optical response of individual silver nanohedra}} \\
		\vspace{0.3cm} \\
		\noindent\large{Yisu Wang\textit{$^{a,\circ}$}, Zoltan Sztranyovszky\textit{$^{b,\circ}$}, Attilio Zilli\textit{$^{a,c}$}, Wiebke Albrecht\textit{$^{d,\ddag}$}, Sara Bals\textit{$^{d}$}, Paola Borri\textit{$^{a}$}, and Wolfgang Langbein\textit{$^{b,\ast}$}} \\
		\vspace{0.3cm} \\
		\noindent\normalsize{
	The optical response of metal nanoparticles is governed by plasmonic resonances, which are dictated by the particle morphology. A thorough understanding of the link between morphology and optical response requires quantitatively measuring optical and structural properties of the same particle. Here we present such a study, correlating electron tomography and optical micro-spectroscopy. The optical measurements determine the scattering and absorption cross-section spectra in absolute units, and electron tomography determines the 3D morphology. Numerical simulations of the spectra for the individual particle geometry, and the specific optical setup used, allow for a quantitative comparison including the cross-section magnitude. Silver nanoparticles produced by photochemically driven colloidal synthesis, including decahedra, tetrahedra and bi-tetrahedra are investigated.  A mismatch of measured and simulated spectra is found when assuming pure silver particles, which is resolved by the presence of a few atomic layers of tarnish on the surface, not evident in electron tomography. The presented method tightens the link between particle morphology and optical response, supporting the predictive design of plasmonic nanomaterials. } 

	\end{tabular}

\end{@twocolumnfalse} \vspace{0.6cm}
]%
%
\renewcommand*\rmdefault{bch}\normalfont\upshape
\rmfamily
\section*{}
\vspace{-1cm}
%

%
\section{Introduction}
Plasmonic \acp{np} have optical properties which are controlled by their morphology. This enables a wide tuneability using a single material, such as silver or gold, just by size and shape control,\cite{TruglerBook16} including chirality and the associated chiro-optical response.\cite{LeeNS20}
The \ac{np} optical properties are described in terms of the cross sections for optical scattering (\sig{sca}{}) and absorption (\sig{abs}{}), which represent the strength of the \ac{np}--radiation interaction.\cite{BohrenBook98}
While many experimental techniques have been developed to characterize the optical response at a single \ac{np} level,\cite{CrutCSR14,OlsonCSR15} only few of these methods are able to quantify both optical cross-sections in absolute units,\cite{HusnikPRL12} or equivalently, the complex polarizability of the \ac{np}.\cite{KhadirO20}

Previous studies of correlative single NP optical--electron microscopy using scattering spectra show the complex and sensitive dependence of the optical response on the morphology.\cite{HenryJPCC11} Numerical modelling of the optical response based on a 3D reconstruction from electron tomography was shown in \Onlinecite{PerassiNL10}, using \ac{dda} simulations of a faceted gold \ac{np}, and for large irregular gold \acp{np} simulated scattering spectra were compared with experiments.\cite{PerassiACSN14} Furthermore, gold--silver core--shell \acp{np} were investigated, either showing simulations for a given morphology \cite{HernandezNS14} or comparing simulations with measured scattering spectra as function of shell thickness.\cite{ChuntonovNL11}
However, the above works did not attempt an accurate comparison of measured and simulated cross-sections, and focussed on the spectral features instead. Over the past years, we have developed a measurement and data analysis method to retrieve accurate quantitative cross-section spectra.\cite{ZilliACSP19,BorriUSpatent21} In \Onlinecite{WangNSA20} we combined this method with standard projection  \ac{tem} to investigate silver cubes. The cube geometry results in \acp{np} orientated such that one of the flat sides is attached to the \ac{tem} grid, so that the \ac{np} geometric parameters can be reasonably extracted from projection images. For more complex shapes, however, conventional \ac{tem} is insufficient, and electron tomography is needed. 

In the present work, we study faceted silver \acp{np}  produced by photochemically driven colloidal synthesis,\cite{PietrobonCM08,ZhengL09,YeNS15} including decahedra, tetrahedra and bi-tetrahedra.
Similar to their gold counterparts,\cite{DasJPCC12,MyroshnychenkoNL12} their response is ruled by \aclp{lspr}.
The chemical reactivity of silver surfaces makes these systems attractive for catalysis applications,\cite{AlekseevaACSP19,LiaoNS19} but also provides a route to chemical surface modifications which can be difficult to identify in \ac{tem} images while significantly modifying the optical response.\cite{KangCR18} We find here that an accurate quantitative study of cross-section spectra correlating experiment with simulation can uncover such detail. 
The presented case study on the one hand assesses the level of accuracy that can be achieved by our cross-section measurement method, and on the other hand exemplifies the kind of fine information that can be extracted from quantitative cross-section spectroscopy. Ultimately such progress might enable to reliably extract the 3D morphology of metal NPs from optical measurements alone.

\begin{figure*}[htb]
	\includegraphics*[width=\textwidth]{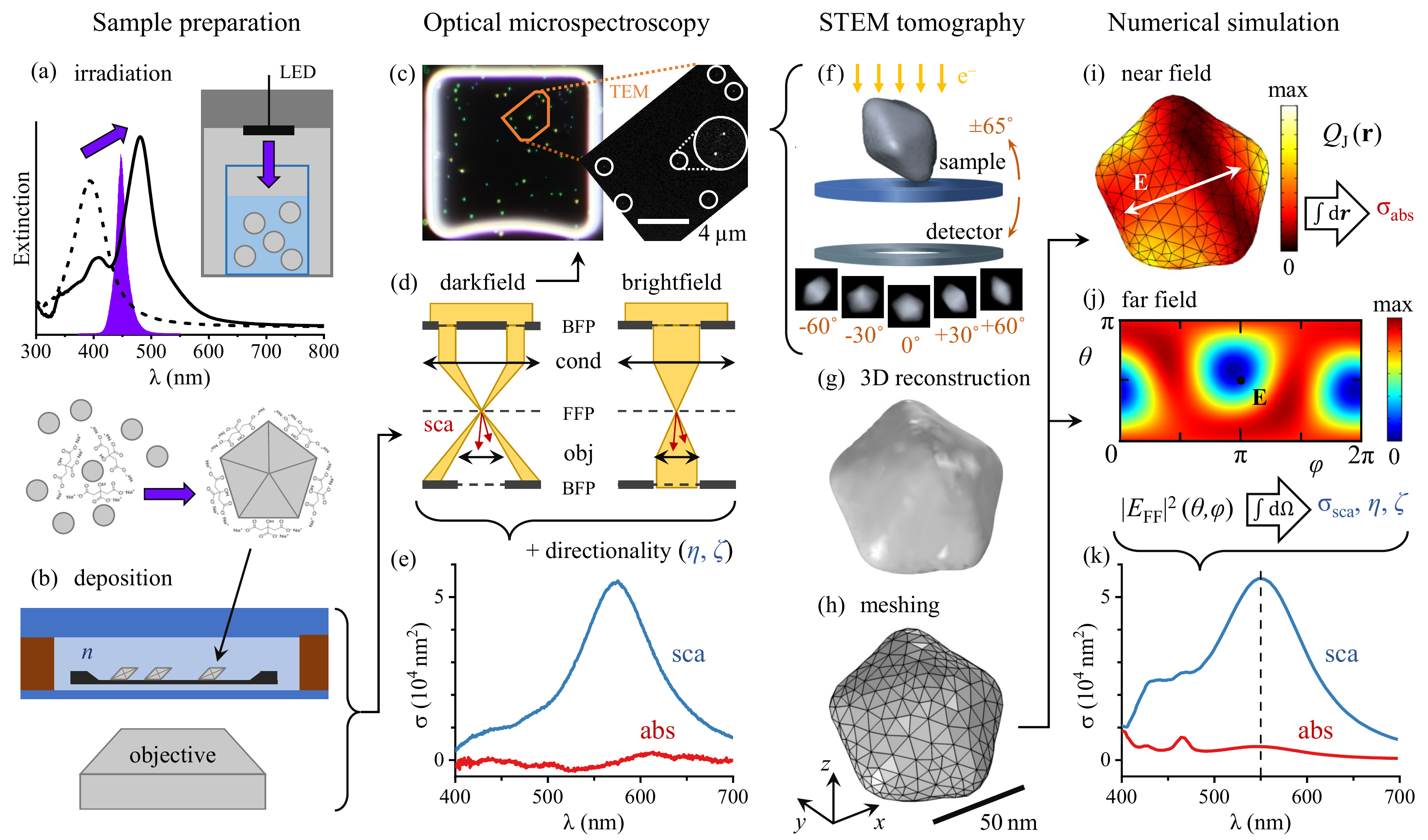}
	\caption{\label{fig:workflow}
		Schematic workflow as described in the text. a) Photochemical formation
		of decahedra using blue LED illumination, monitored via the red-shift
		of the extinction from spherical seeds (dashed line) to decahedra (solid line). b) Deposition of decahedra onto a TEM grid with SiO$_2$ windows, index-matched by anisole  immersion, and encapsulated by a glass slide and a coverslip. c,d) Optical micro-spectroscopy in dark-field and bright-field configurations.  BFP and FFP indicate, respectively, the back and front focal plane of the objective (obj) and condenser (cond) lens. e) Measured single-decahedra scattering and absorption cross-section spectra in absolute units. f) Correlative HAADF-STEM tomography through recognition of NP patterns as exemplified in c). g) 3D shape reconstruction from tomography. h) Tetrahedral volume mesh used in numerical simulations. i) Calculated spatial distribution of the Joule (resistive) heating. j) Calculated far-field distribution of the scattering intensity. k) Numerical simulations of cross-section spectra. Panels e--k refer to the exemplary particle \#20.		
	}
\end{figure*}
\section{Materials and methods\label{sec:experiment}}
Let us present the workflow of the experiment summarized in \autoref{fig:workflow}.
Silver decahedra \acp{np} are fabricated with a plasmon-driven method adapting the protocols of \citeauthor{ZhengL09}\cite{ZhengL09} and \citeauthor{PietrobonCM08}\cite{PietrobonCM08}.
As shown in \autoref{fig:workflow}a, seeds grown by reduction of AgNO\tb{3} in aqueous solution are thought to aggregate to form decahedra under irradiation by a high power \ac{led} centred at a \SI{447}{\nm} wavelength (violet spectrum in the graph). The formation of decahedra can be monitored via the progressive red-shift of the extinction peak of the \ac{np} solution from spherical seeds (dashed line) to decahedra (solid line). Further details of the fabrication process and a kinetic study are reported in the ESI \autoref{S-sec:sample_fabrication}.

As particle support for the correlative measurements we used a \ac{tem} grid (Ted Pella, 21530-10) composed of a \SI{40}{\nm}-thick SiO\tb{2} film (refractive index $n = \num{1.46}$) supported by a \SI{200}{\nm}-thick Si\tb{3}N\tb{4} film with $50\times50$ \textmu m square windows, on a silicon substrate (one such window is the bright frame of \autoref{fig:workflow}c). The grid was washed using two repetitions of the sequence deionised water -- acetone -- anisole -- ethanol, and then dried in air. The grid was held by a teflon-coated stainless steel reverse-action tweezer throughout the functionalisation and washing process. The grid was incubated for 1 hour at 55\,$^\circ$C in 10\,mL etching solution of 500\,\textmu L HCl (99\%) diluted in 9.5\,mL of 30\% H$_2$O$_2$. The grid was then washed three times in water, followed by three times in ethanol. 200\,\textmu L of (3-Aminopropyl) triethoxysilane (APTES) (Sigma Aldrich) was centrifuged at 20k RCF for 20\,mins to spin down any large debris. 100\,\textmu L of this APTES stock was then diluted in 9.9 mL ethanol (absolute, for HPLC, >99.8\%, Sigma Aldrich) to obtain a 1\% APTES solution, in which the grid was incubated for 1 hour. The grid was then washed three times in ethanol followed by three times in water. The resulting functionalised grid was dried in air at 55\,$^\circ$C for 30 mins and stored at 4\,$^\circ$C for no longer than one month.
The decahedra solution (9\,\textmu l of 0.25 optical density at 475\,nm) was wet-cast (see \Onlinecite{WangNSA20}) onto the functionalised grid. The grid was subsequently washed by gently and repeatedly dipping in water, and then dipped in ethanol and dried. 

To provide the NPs with a nearly homogeneous optical environment for the cross-section measurements, the \ac{tem} grid was sealed in anisole ($n = \num{1.52}$) between a microscope slide ($25 \times 75$\,mm$^2$, Menzel Gl\"aser) and a coverslip (\#1.5, $25\times25$\,mm$^2$, Menzel Gl\"aser) using a \SI{0.5}{\mm} thick adhesive silicone spacer (Grace Bio-Labs 664507), with the TEM grid surface facing the coverlip side. We chose anisole rather than microscope immersion oil as it is volatile and evaporates without leaving residuals, enabling subsequent electron microscopy. This assembly is mounted onto an inverted optical microscope (Nikon, Eclipse Ti-U) with a 40$\times$ dry objective (Nikon MRD00405, CFI plan apochromat $\lambda$
series) of 0.95 \ac{na} as depicted in \autoref{fig:workflow}b.

The procedure for the optical measurements and the quantitative analysis of the optical cross sections is largely the same we adopted in \Onlinecite{WangNSA20}.
We therefore limit ourselves here to recapitulate the main steps performed and parameters used, while we refer the reader to the aforementioned work\cite{WangNSA20} for an in-depth description.
Single-particle microspectroscopy is performed by optically relaying the intermediate image plane created by the tube lens of the microscope onto the entrance slit of an imaging spectrometer (Horiba Jobin-Yvon, iHR550) equipped with a ruled plane diffraction grating (Horiba, 51048) of \SI{78}{\mm} square size and \num{100} lines per mm.
Spectra were acquired with a Peltier-cooled back-illuminated \ac{ccd} sensor (Andor, Newton DU-971N).
The spectrometer images the entrance slit onto the sensor, allowing to use the zeroth order of the grating to provide an image of the sample to select a specific particle for spectroscopy.
The entrance slit acts as a spatial filter in the horizontal direction (along the spectral dispersion), whereas in the vertical direction the binning of the \ac{ccd} sensor itself is used to define a region of interest.
Together these define a $1.0 \times 1.0$\,\textmu m square region centred on the \ac{np} of interest from which the signal is collected. The corrections required to account for this finite region of detection are described in \autoref{S-sec:finite_detection} of the ESI.

Within the transillumination scheme adopted, we define two imaging modalities based on the angular range of the illumination, as illustrated in \autoref{fig:workflow}d.
In the first one -- a \ac{bf} scheme -- the illumination \ac{na} range is set to match the collection range (0--0.95) of the objective.
In the second one -- a \ac{df}  scheme, the illumination range 1.06--1.34\,\ac{na} is used, not overlapping with the collection range, so that only scattering is detected.
As a result, scatterers such as \acp{np} are visible as bright diffraction-limited spots on a dark background -- see for example \autoref{fig:workflow}c (left).
The two illumination ranges are defined by two corresponding 3D-printed apertures placed in the \ac{bfp} of the condenser lens (Nikon, T-C-HNAO, \SI{1.34}{NA} oil-immersion) on a slider, which allows the reproducible switching between \ac{bf} and \ac{df} required for an accurate correlation between transmitted and scattered light intensity.

The optical cross sections are defined as the power removed from the exciting beam per excitation intensity: $\sigma = P/I_\mathrm{exc}$.
Thus, a careful referencing to the exciting intensity\cite{PayneSPIE19} of the single-particle extinction and scattering spectra enables us to measure accurately the magnitude of the cross sections.
Note that the \ac{bf} extinction signal includes contributions of both absorption and scattering, which have to be unravelled based on the scattering-only \ac{df} signal.
Such retrieval procedure is presented in \Onlinecite{ZilliACSP19}, and requires information on the directional properties of the scattering process.
In the analysis this information is reduced to two parameters named $\eta$ and $\zeta$.
$\eta$ concerns the detection, and is the fraction of the total scattering collected by the objective.
We note that $\eta$ depends on the angular range of the illumination, such that $\eta^\textsc{bf}\neq \eta^\textsc{df}$; however, the  difference is small for the decahedra, whose response is governed by the same dipolar mode under both \ac{bf} and \ac{df} illumination.
$\zeta$ concerns the excitation, and is the \ac{bf}-to-\ac{df} ratio of the scattered power; it depends therefore on the relative intensity of the \ac{bf}-to-\ac{df} illumination (which we characterised for our set-up as described in \autoref{S-sec:xi} of the ESI), as well as on how much the resonant modes of the scatterer are excited under either illumination.
In this work, $\eta$ and $\zeta$ are computed numerically for each studied \ac{np} as described below. The details of the \ac{np} geometry for the cases studied here have a moderate effect, and therefore the values are rather similar for all \acp{np} considered, see the ESI \autoref{S-sec:all_spectra}.
Following the quantitation procedure outlined above, we can measure cross-section spectra in absolute units, such as \SI{}{\nm^2} in \autoref{fig:workflow}e.
Note that $\sig{sca}{}(\lambda)$ and $\sig{abs}{}(\lambda)$ refer to a given illumination and collection range. Specifically, in this work we measure \sig{sca}{df} and \sig{abs}{bf}, which differ\cite{PayneSPIE19} from the cross sections under plane-wave excitation. 

As illustrated by \autoref{fig:workflow}c, optical and electron microscopy images can be correlated through the recognition of a specific \ac{np} pattern. In the high-angle annular dark-field scanning TEM (HAADF-STEM) overview on the right, white circles highlight the \acp{np} visible, and a distinctive dimer in the middle is shown magnified.
We are thereby able to select the \acp{np} characterised optically for HAADF-STEM tomography, wherein the sample is tilted across a wide angular range under the electron beam, as depicted in \autoref{fig:workflow}f, and the resulting stack of projection images is used to reconstruct the three-dimensional (3D) morphology of the \ac{np}.
All electron tomography series were acquired using a FEI Tecnai Osiris electron microscope operated at \SI{200}{\kV}.
The series are taken across the largest tilt range allowed by the \ac{tem} grids clearance -- typically about \SI{\pm 65}{\degree} -- with a tilt increment of \SI{3}{\degree}\!.
The 1k\,$\times$\,1k projection images are aligned to match the NP positions across each series using cross-correlation, and are then reconstructed using 15 iterations of the expectation--maximization reconstruction algorithm implemented in the ASTRA toolbox for \matlab.\cite{MoonSPM96,AarleU15}
The resulting reconstructions are downsampled by a factor 12 and segmented using the Otsu method to export them as .stl files, such as the one shown in \autoref{fig:workflow}g.
This geometry is then meshed in \textsc{comsol} for numerical simulation purposes with a free tetrahedral volume mesh displayed in \autoref{fig:workflow}h. The influence of variations of this reconstruction procedure on the simulated cross-section spectra is investigated in \autoref{ss:geometry}. 

The optical response of the particles is computed in the frequency domain using \textsc{comsol} Multiphysics\textsuperscript\textregistered\!, a commercial software implementing the finite-element method.
In the model, the \ac{np} is defined as silver using the permittivity reported in \Onlinecite{YangPRB15}, immersed in a homogeneous medium of anisole ($n = \num{1.52}$). 
We neglected the small index mismatch between the thin silica window ($n=1.46$) and anisole and used a homogeneous medium instead of a multi-layered structure, therefore the model used here is equivalent to the one described in the SI of our previous work \Onlinecite{WangNSA20}, with the slab thickness set to zero ($d=0$\,nm). This simplification allowed us to automate the importing and alignment of particle geometries from HAADF-STEM tomography into \textsc{comsol}. The stationary solution of Maxwell's equations under  \ac{pw} excitation of given frequency, polarization, and propagation direction computed by \textsc{comsol} determines the spatial distribution of the electromagnetic field $\vec{E}$.

Let us now discuss how we derive the observables of interest (namely \sig{abs}{}, \sig{sca}{}, $\eta$, $\zeta$) from this solution.
\autoref{fig:workflow}i shows the spatial distribution of the Joule (resisitive) heating $Q_\textsc{j} = \frac{1}{2} \Re (\vec{J}_\mathrm{c} \cdot \vec{E}^*)$ where $\vec{J}_\mathrm{c} = \varsigma \vec{E}$ is the conduction current in terms of the AC electrical conductivity $\varsigma$.
We integrate $Q_\mathsc{j}$ over the \ac{np} volume to compute the absorbed power $P_\mathrm{abs}$, and hence $\sig{abs}{pw} = P_\mathrm{abs}/I_\mathrm{exc}$ dividing by the excitation intensity $I_\mathrm{exc}$.
The near-field solution can be projected to the far field via the far-field transform available in \comsol, resulting in an angular distribution of the field $\vec{E}_\mathsc{ff}(\theta,\phi)$  such as the one shown in \autoref{fig:workflow}j.
A dipole-like emission pattern is seen, with the dipole oriented close to the $x$ direction (identified by the polar angle $\theta = \pi/2$ and the azimuth $\phi = 0,\pi,2\pi$) -- albeit not precisely along it, due to a tilt of a long axis of the particle, along which its polarizability is maximized.
The far-field Poynting vector $\vec{S}_\mathsc{ff}$ (which is proportional to $\abs{\vec{E}_\mathsc{ff}}^2$ plotted in \autoref{fig:workflow}j) can be integrated over the appropriate solid angle ($4\pi$ or the objective acceptance) to compute the scattered power $P_\mathrm{sca}$ (and hence $\sig{sca}{pw} = P_\mathrm{sca}/I_\mathrm{exc}$) and the collected fraction of scattering $\eta$.

We emphasize that these values of $\sigma$ and $\eta$ are computed under \ac{pw} excitation, which we have indicated with the \acs{pw} superscript; in the experiment instead we use the incoherent illumination produced by a high-\ac{na} condenser, which is composed of a wide range of directions.
To reproduce the measured \sig{sca}{df} and \sig{abs}{bf} we therefore perform and average a large number of \ac{pw} simulations sampling the directional range of illumination (either \ac{bf} or \ac{df}), each direction being assigned an appropriate weight according to the angular dependence of the illumination intensity in our microscope, that we have characterised.
The unpolarised illumination is reproduced by averaging for each direction the results of two \ac{pw} simulations with orthogonal excitation polarisation, namely p and s with respect to the imaged sample plane.
An analogous directional averaging is applied to compute the scattering parameters ($\eta^\textsc{bf}$, $\eta^\textsc{df}$, and $\zeta$) appropriate to the experimental angular ranges of illumination and detection.
The mathematical details of such procedure are given in \Onlinecite{WangNSA20} section S.IV to S.VI.	

This averaging results in the \sig{sca}{df} and \sig{abs}{bf} spectra, which are shown in \autoref{fig:workflow}k, and are quantitatively simulating the experimental ones in \autoref{fig:workflow}e.
From here on we drop the \ac{df} and \ac{bf} superscript of the cross-sections for simplicity.  
In the next section we will compare in detail the experimental and simulated cross-sections, focussing on their differences to identify additional aspects of the system beyond its measured geometry yet to be included in the model. In this manner, the comparison can bring about additional knowledge on the system -- such as the presence of surface layers or variations of the metal permittivity.
\section{Results and discussion}
\label{sec:results}
\begin{figure*}[htbp]
	\includegraphics*[width=\textwidth]{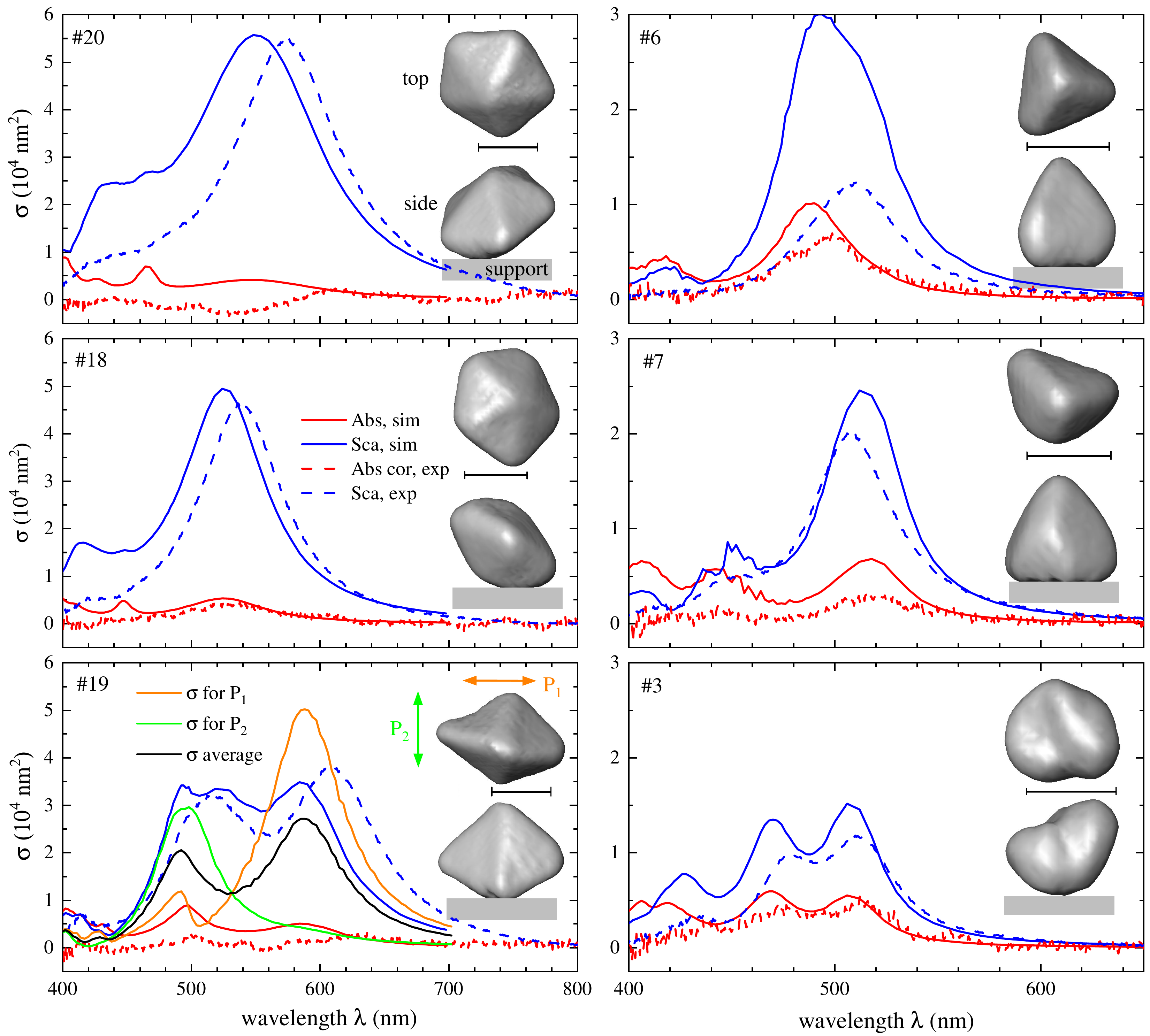}
	\caption{\label{fig:fig2}
		Measured (dashed lines) and simulated (solid lines) scattering (blue) and absorption (red) cross-section spectra of 6 selected particles as labelled, along with HAADF-STEM tomography surface views from the top and side. The scale bar is 40\,nm. For particle \#19, we show additionally the simulated scattering cross-section for normal incidence for linear polarizations along (orange line) and across (green line) the long axis of the particle, as well as their average (black line).
	}
\end{figure*}
\begin{figure}[htbp]
	\includegraphics[width=\columnwidth]{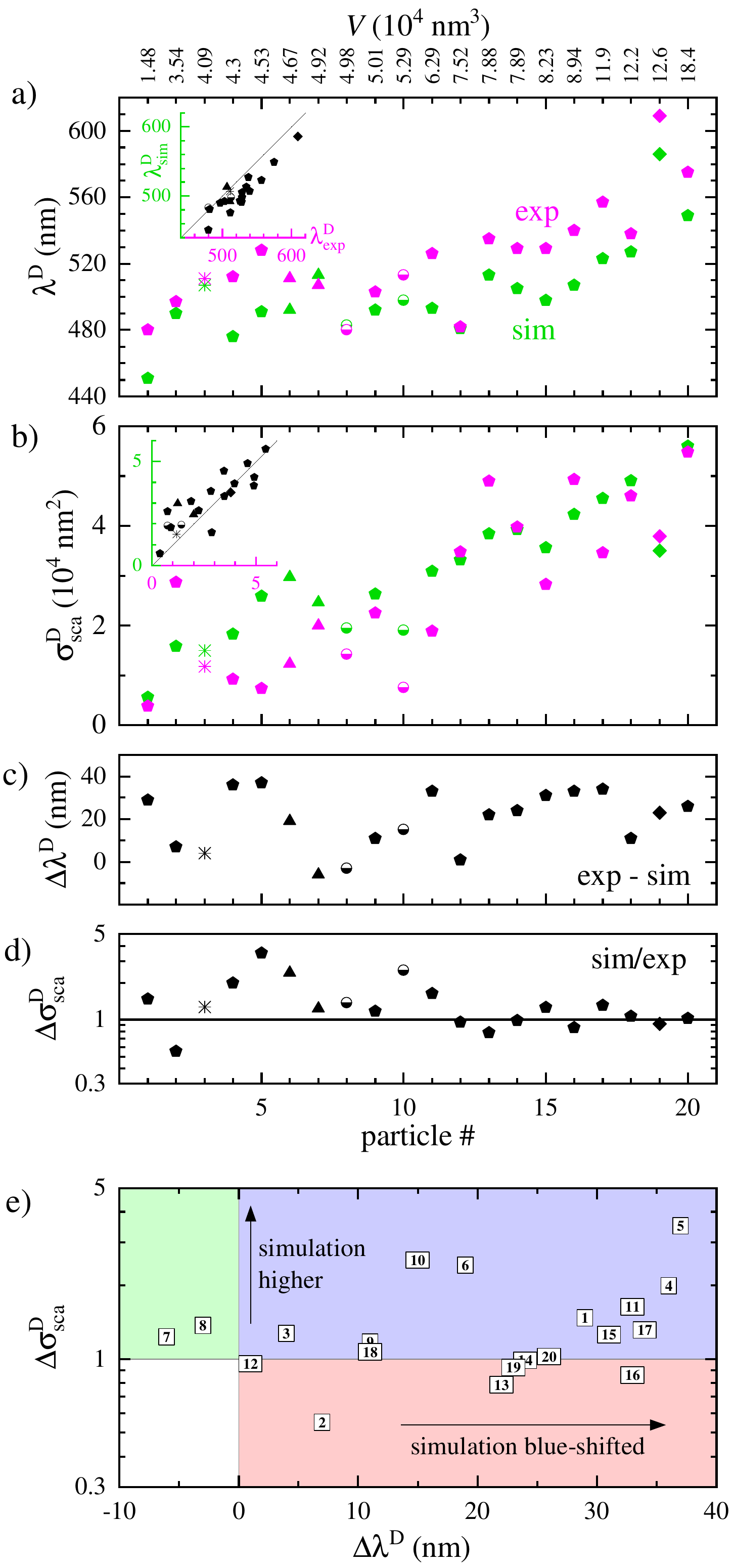}
	\caption{\label{fig:fig3}
		Comparison of measured and simulated properties of the dipole peak in the scattering cross-section spectra for all investigated particles. a) Position of the peak $\lambda^{\rm D}_{\rm sca}$. For particles with multiple peaks, such as \#19 or \#3, the longer wavelength peak is shown. The symbols are indicative of the particle shape (see insets in \autoref{fig:fig2} and ESI \autoref{S-sec:all_spectra}): \#6 \& \#7 are tetrahedra, \#8 \& \#10 are half spheres, 19 is a bitetrahedron, \#3 is not well defined, the rest are decahedra. The inset shows simulated versus measured positions. b) Amplitude of the peak. The inset shows simulated versus measured amplitudes. c) Difference between the simulated and experimental peak position. d) Ratio between simulated and experimental peak amplitude. e) Peak amplitude ratio versus position difference. 
	}
\end{figure}
Twenty particles were measured in total, which we numbered with increasing volume $V$. \autoref{fig:fig2} shows the measured and simulated cross-section spectra for six selected particles representing the range of shapes and sizes, along with the top and side view of their 3D reconstructions. The data for the remainder of the particles are shown in the ESI \autoref{S-sec:all_spectra}. Animated 3D renderings of the NP reconstructions are shown in the ESI \autoref{S-subsec:3Dmovies}. The top view shows the particle as seen along the illumination axis, indicating the main plane of excitation polarizations, even though due to the high NA also axial polarization is present, more markedly for the \ac{df} illumination.  While the fabrication method was developed to produce decahedra (such as particles \#20 and \#18), other shapes are present, such as tetrahedra (\#6 and \#7), or a bi-tetrahedron (\#19). The particles range in sizes, as summarised in the ESI \autoref{fig:fig3}. The decahedra and tetrahedra show a single pronounced peak in the scattering cross-section, at a wavelength between 500 and 550\,nm. The more elongated particles, \#19 and \#3, show two distinct peaks, which are dipolar modes with polarisations along the longer or shorter axis, centred at longer or shorter wavelengths, respectively. \textsc{comsol} simulations of the scattering cross section of particle \#19 under normal-incidence plane-wave illumination polarized along the shorter and longer axis (green and orange lines, respectively) confirm this attribution. For most particles we find a reasonable agreement in the lineshape and magnitude of the scattering cross-section peak around the dipolar resonance, though the position shows a systematic blue-shift of the simulated data relative to the measured one. The measured absorption spectra show regions of negative values, which is not expected, as it implies a net power emission by the particle. The absorption is determined as difference between extinction and scattering, using a range of numerically calculated and experimentally measured parameters, as mentioned in \autoref{sec:experiment}. With the scattering dominating for most particles, the resulting small difference is affected by systematic errors in the measured extinction and scattering. These considerations and the wavelength dependence of the analysis parameters are discussed in more detail in the ESI \autoref{S-sec:all_spectra}. 

To correlate the results of experiment and simulations across all particles measured, we compare key spectral features in \autoref{fig:fig3}. The position of the dipolar scattering peak (panel a) shows a redshift with increasing particle index and thus particle volume, and the amplitude of the peak (b) increases with volume, both of these effects are generally well known and understood in literature.\cite{ZhangACIE09,YeNS15} The quantitative comparison between measurements and simulations shows a remarkable agreement, considering that no adjustable parameters have been used. The difference between simulated and measured peak positions can be seen in the inset, and separately in (c). We find good correlation, with most particles showing a red shift of the measurement relative to the simulation by a few tens of nanometres. This finding is reminiscent of the shift observed in experiments with silver cubes.\cite{WangNSA20} The relative deviation between simulated and measured peak magnitude (see panel d) shows a significant fluctuation, mostly with the simulation being higher, though the deviation decreases for large particles. Generally, the signal-to-noise ratio in the HAADF-STEM projection images is smaller for smaller particles, allowing for a larger relative error.  In addition, the finite angular range used for the tomography reconstructions gives rise to a so-called missing wedge artefact, a result of a lack of information along certain directions. This can lead to systematic errors depending on the particle morphology, which could cause particle to particle fluctuations. On the optical measurement side, smaller cross-sections are more affected by noise due to diffuse background scattering. However, the noise level is typically not significant in the present data, as can be seen in the scattering spectra shown in \autoref{fig:fig2}. On the other hand, the absorption displays a better agreement for small particles, as can be seen in the ESI \autoref{S-sec:all_spectra}. This is due to the response of large particles being dominated by scattering, and the systematic error in the quantification of the absorption being proportional to the scattering, as previously discussed. 

In \autoref{fig:fig3}e the particles are shown in a plane spanned by the ratio in amplitude and the difference in peak position between the measured and simulated data, to facilitate identifying and categorizing the possible sources for the discrepancy. The area shaded in red corresponds to both a blue shift and a decrease in amplitude of the simulated scattering dipole peak compared to the experimental one. It is known that rounding the edges of the particle causes a blue shift and a decrease of the magnitude of the plasmonic peaks. For example, it was observed for silver prisms,\cite{GrzeskiewiczP14} silver cubes,\cite{WangNSA20, RazimanOE13} and gold decahedra.\cite{RodriguezFernandezJPCC09} We note that the samples were shipped from the optical experiment at Cardiff to the electron tomography at Antwerp in nitrogen atmosphere in a sealed container at room temperature, providing up to 4 days during which such rounding might have developed.\cite{MarksJPCM16} In the area shaded in blue the simulated peak is also blue shifted, but the simulated amplitude is higher than the experimental one. Based on our previous work\cite{WangNSA20} this is likely due to a surface layer forming on the particles. An increased damping in the permittivity can also lead to a decrease of the scattering cross section, as we shall discuss in \autoref{ss:permittivity} below. The green area corresponds to a red shift of the simulated spectra with respect to the experimental ones, with an increase in amplitude. The two particles in this area show rather small deviations, within the accuracy of determining the values. 

Below we investigate some of these potential sources of deviation in more detail on two selected particles. As the cross-section simulations taking into account the wide NA range of the microscope illumination are computationally expensive, we increased the sampling step size of the illumination direction from 0.21\,NA to 0.3\,NA, reducing the simulation time by a factor of two, while affecting the cross-section spectra by less than a few percent.

\subsection{Geometry reconstruction accuracy}
\label{ss:geometry}

The measured NP morphology dictates the simulated optical cross-sections, and thus should be as accurate as possible. In our analysis pipeline, the reconstruction of the electron tomography depends on analysis parameters which influence the resulting morphology. As mentioned earlier, electron tomography suffers from the missing wedge artefact, which leads to a lack of information along certain directions, and we found that the resulting morphology slightly depended on the number of iterations in the reconstruction process. One can also include pre-processing of the data such as smoothing procedures. In addition, to achieve a reasonable simulation time, the NP morphology needs to be meshed with an acceptable number of elements, which depends on the computational power available and the accuracy required. In this section we discuss the influence of these points on the reconstructed morphology and simulated spectra. 

\begin{figure}
	\includegraphics[width=\columnwidth]{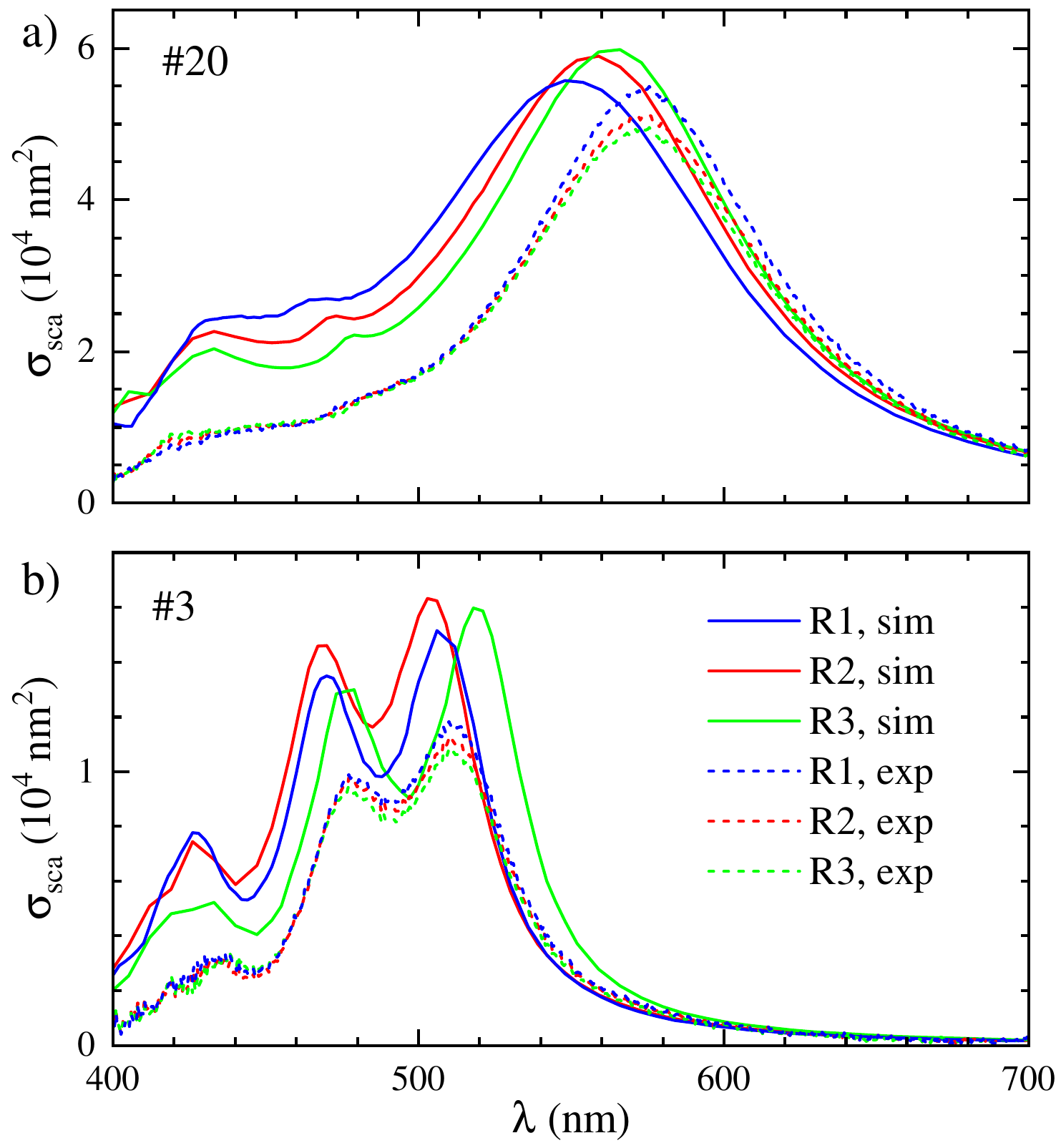}
	\caption{\label{fig:fig4}
		Simulated and measured scattering cross-section spectra for particle \#20 (a) and \#3 (b) using different tomography reconstruction settings R1 to R3 as labelled (see text).
	}
\end{figure}

We call R1 the meshed reconstructions used in \autoref{fig:fig2} and \autoref{fig:fig3}, which employed 15 iterations of the expectation--maximization reconstruction algorithm and a downsampling factor of $N=12$. Downsampling by a factor $N$ bins together pixels in an $N \times N \times N$ volume, so reduces the number of elements defining the NP's surface by a factor of $N^2$. The exact number of facets depended on the NP, but in general for R1 the NP's surface geometry consisted of a few thousand faces. In the reconstruction procedure R2, we smoothed the input projection images with a pixel radius of 3 prior to the iterations to improve the signal-to-noise ratio, and reduced $N$ to 4, which increased the number of surface elements to tens of thousands. For reconstruction R3, we furthermore increased the iterations to 100 and reduced $N$ to 1, which increased the number of surface elements to hundreds of thousands. 

For the large number of surface elements resulting from R2 and R3, \textsc{comsol} was unable to reliably import the geometry and construct usable particle models. To circumvent this problem we reduced the number of surface elements to approximately 1000 before importing. This did not cause a significant loss of accuracy: we observed typically around 5\,nm blue shift and $1\%$ increase in amplitude (see ESI \autoref{S-sec:geometry_modification} for details about the procedure and the effects). We note that the mesh on which \comsol\ solves the scattering problem is usually even coarser. This mesh was determined by investigating the convergence of the simulated scattering cross-section amplitude at the dipole peak versus the mesh size, as described in the ESI of \Onlinecite{WangNSA20} -- we choose the size of NP mesh elements so that the calculated dipole resonance scattering amplitude is within 1\% from the converged value, yielding about 500-1000 surface elements on the NP.

The reconstructions R2 and R3 resulted in a slightly altered geometry that was hardly discernable visually on the \textsc{comsol} mesh, so that we look here at the calculated volume and surface area changes (see ESI \autoref{S-tab:volumes}), and the effect on the cross-section spectra, as shown in \autoref{fig:fig4} (see ESI \autoref{S-fig:geometry} for more examples). For particle \#20, the volume is $V = (18.4,18.2,17.4)\times 10^4$ nm$^3$ for (R1, R2, R3), and the volume to surface ratios are $V/S= (10.8, 10.6, 10.4)$  nm. For particle \#3 the volumes are $V = (4.09,4.29,4.10)\times 10^4$ nm$^3$ and the volume to surface ratios are  $V/S =  (6.65,6.79,6.5)$ nm.

For the larger particle (\#20 shown in \autoref{fig:fig4}a), the reconstructions have little influence on the simulation results. Despite the decreasing volume, we observed a small red-shift and increase in scattering cross-section for R2 and R3. Noting that these reconstructions create less smoothing of morphological features, the red shift can be related to a sharpening of the geometry. For the smaller particle (\#3 shown in \autoref{fig:fig4}b), R2 and R3 create different effects. For R2 we observe a small blue shift and a small increase in amplitude. The blue shift could result from remeshing, as mentioned before. The increase of the scattering amplitude is consistent with the increase in the volume. For R3 instead, we observe a red shift and slight redistribution of amplitude between the two peaks is seen. We attribute this to a sharpening of morphological features in the missing wedge region due to the higher number of iterations in the reconstruction algorithm. The slight increase in the splitting of the two peaks also suggests a small increase in aspect ratio. The modified simulated cross-sections result in modified analysis parameters ($\eta$, $\zeta$) which in turn modify the measured cross-sections slightly, as shown by the dashed lines. 

The results discussed in this section are indicative of the uncertainty originating from the reconstruction. For the following simulations we chose to use R2, having a slightly improved signal-to-noise ratio compared to R1 due to the additional smoothing of the input projections, but avoiding R3 where the high number of iterations may lead to a roughening of the morphology by an overfitting of noise in the expectation--maximization algorithm.

\subsection{Modification of the permittivity}
\label{ss:permittivity}
It is well known that the permittivity of a metal measured by ellipsometry on a planar surface can require a modification for \acp{np} due to the reduced mean free path of the electrons. \cite{VoisinJPCB01}
We accordingly model the effect of additional damping (combining the surface damping, the so-called chemical interface damping, and crystal defects) on the Ag permittivity $\varepsilon_{\rm exp}(\omega)$ measured by ellipsometry on a planar surface of polycrystalline Ag films \cite{YangPRB15} as function of the angular frequency $\omega = 2\pi c/ \lambda$, with the speed of light $c$ and the wavelength $\lambda$.  We first fit $\varepsilon_\mathrm{exp}(\omega)$ in the wavelength range between 400\,nm and 700\,nm, avoiding the Ag inter\-band transitions at shorter wavelengths, with a Drude model, $\varepsilon(\omega,\gamma) = \varepsilon_\infty - \omega_{\rm p}^2 / (\omega^2 + i \omega \gamma)$, as detailed in the ESI \autoref{S-sec:permittivity_modification}, where $\omega_{\rm p}$ is the plasma frequency and $\gamma$ is the damping. Then, we increase the damping by the term\cite{BerciaudNL05, MasiaPRB12, VoisinJPCB01, MuskensPRB08} $g \vF/R$ , where  $\vF$ is the Fermi velocity, $R$ is the effective radius, and $g$ is a scaling factor. We use the radius $R$ calculated from the particle volume $V$ assuming a spherical shape, $R = \sqrt[3]{3V /(4\pi)}$, resulting in $R=35.2$\,nm for particle \#20 and $R =21.7$\,nm for particle \#3. Finally, we add the permittivity change due to the increased damping to the measured permittivity data set, resulting in the modified permittivity $\varepsilon_{\rm m} (\omega) = \varepsilon_\mathrm{exp} (\omega)+ \varepsilon(\omega,\gamma+g \vF/R)-\varepsilon(\omega,\gamma)$ to be used in the simulation.

\begin{figure}
	\includegraphics[width=\columnwidth]{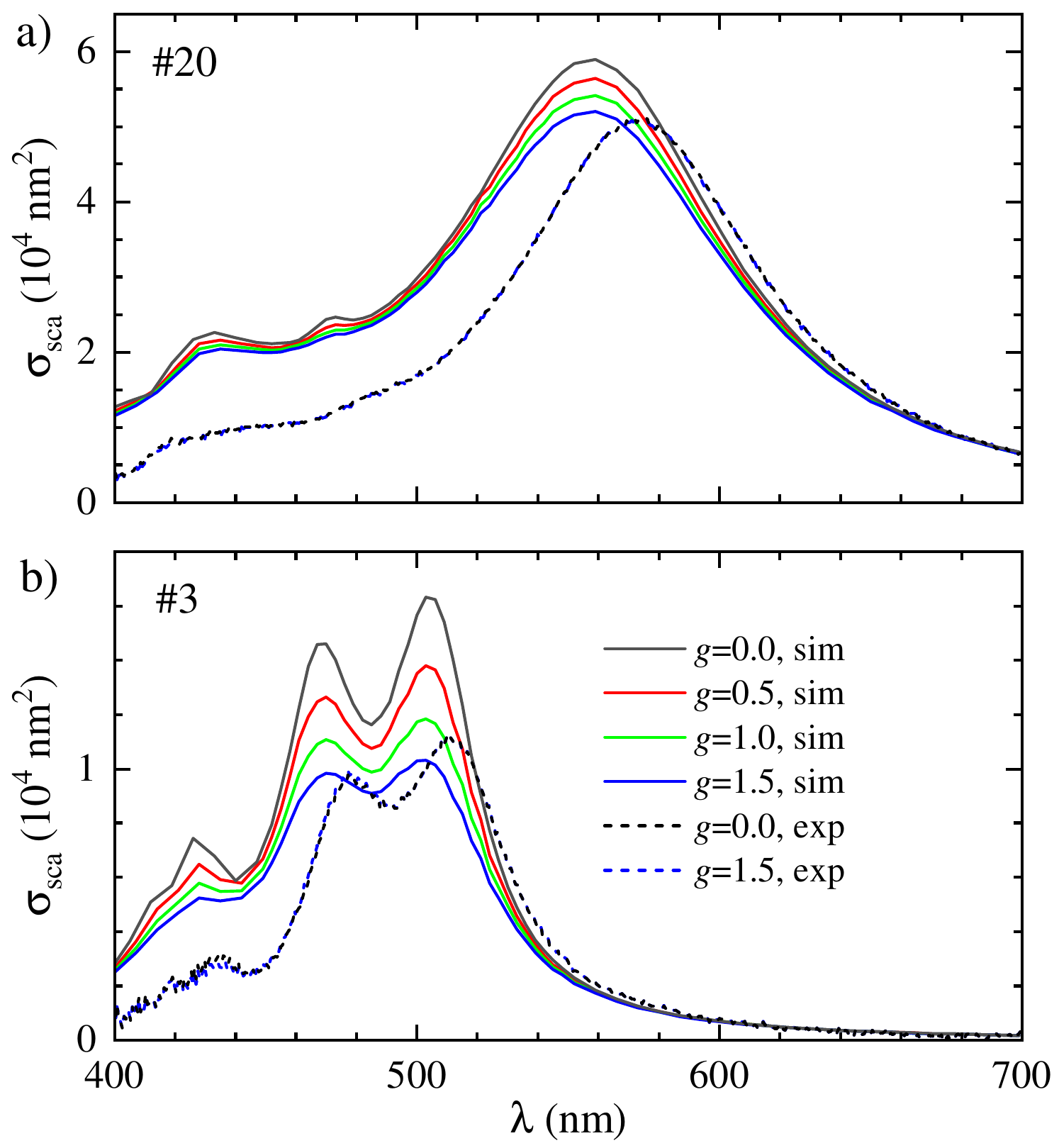}
	\caption{\label{fig:fig5}
		Same as \autoref{fig:fig4}, but for increasing surface scattering $g \vF/R$ in the Drude damping of the Ag permittivity.
	}
\end{figure}

The effect of the increased damping on the cross-section spectra is shown in \autoref{fig:fig5}. The scattering cross-section decreases with increasing $g$ from 0 to 1.5 (a typical range reported previously\cite{MuskensPRB08}), together with a broadening of the peaks, while the absorption cross-section increases (see ESI \autoref{S-fig:absorption}). The measured cross section does not change notably with $g$, showing that the analysis parameters ($\eta^\textsc{bf}$, $\eta^\textsc{df}$, $\zeta$) are not significantly affected by the additional damping.

\subsection{Addition of a tarnish layer}
\label{ss:tarnish}

\begin{figure}
	\includegraphics[width=\columnwidth]{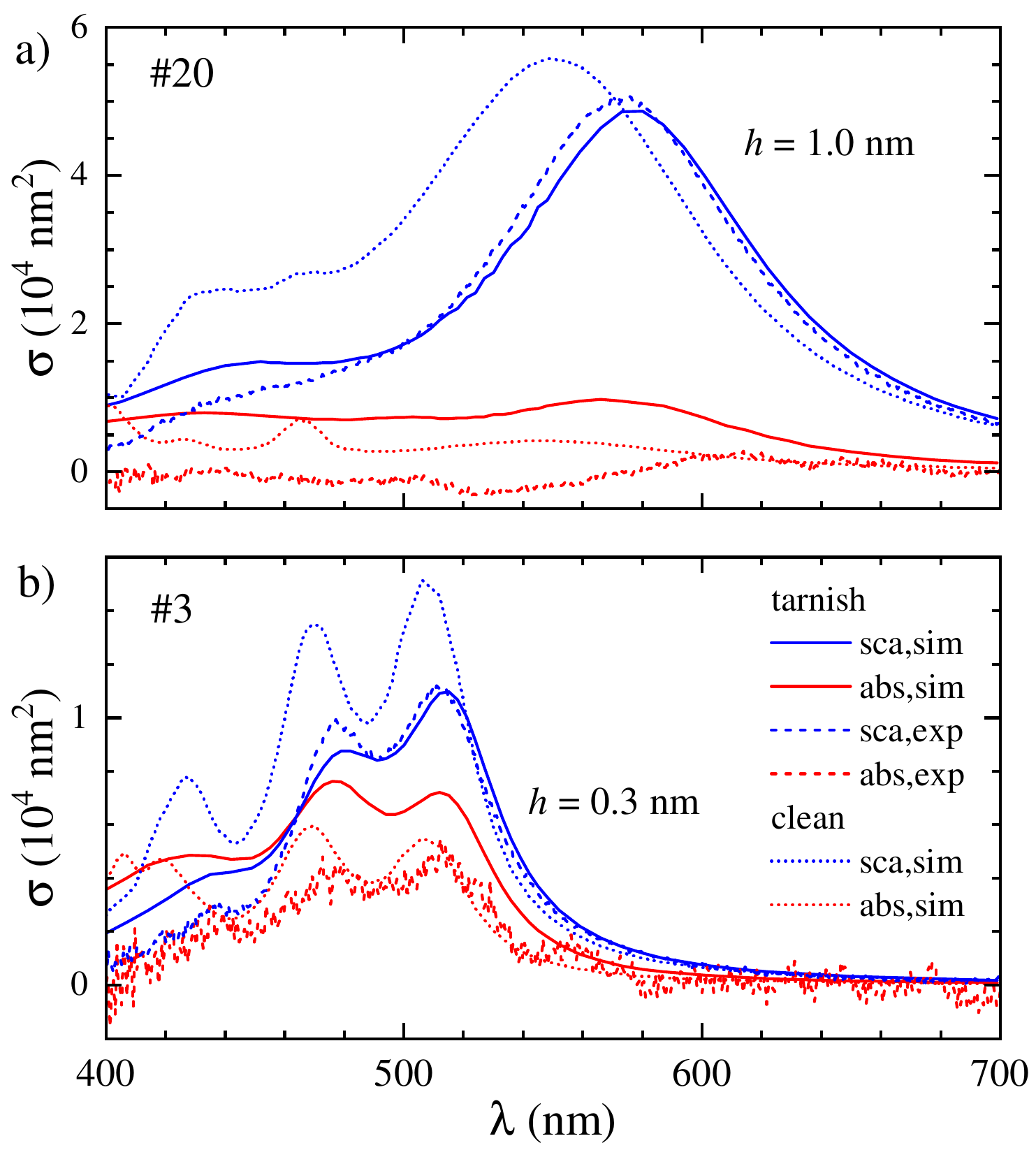}
	\caption{\label{fig:fig6}
		Same as \autoref{fig:fig4}, but for the addition of a silver sulfide (Ag$_2$S) tarnish layer of thickness $h$, and additionally showing the absorption cross-section.
	}
\end{figure}

While it might be possible that changing both reconstruction procedure and damping could produce a $\sigma_\textrm{sca}$ matching the measurements, we did not find a reconstruction that would consistently move the spectra of all particles enough that a further permittivity change could explain the remaining discrepancy. Therefore we consider here another deviation of the particle description in the model from reality, given by an atomically thin chemical surface modification, which is not expected to be visible in the electron tomography for the imaging settings used. Such a  layer, which can form on silver (as opposed to gold) due to its reactivity, is most likely sulfide or oxide.\cite{McmahonAPB05,ElechiguerraCM05} Both compounds have a high refractive index and also absorption, causing a red shift and a decrease in scattering magnitude,\cite{WangNSA20,GrilletJPCC13} mimicking the observed mismatch between simulation and measurement for the majority of the particles.  

More discussion and data regarding  the possible origin and experimental evidence and of such layers, including energy-dispersive X-ray spectroscopy, is given in the ESI {\autoref{S-sec:sulfidisation}}.

To model such layers in \comsol\ we used the following approach: Starting from the surface mesh of the particle, we modeled a surface layer by isotropically scaling down the mesh, while fixing its center of mass, to define a Ag core of volume $V_\textrm{c}$, with the remaining space in the original volume $V_\textrm{s}$ providing the shell. The resulting average shell thickness $h$ is taken as $h = 2(V_\textrm{s} - V_\textrm{c})/(A_\textrm{s} + A_\textrm{c})$, , where $A_\textrm{s}$ and $A_\textrm{c}$ are the surface areas before and after the scaling, respectively. Since sulfur is typically more reactive with Ag than oxygen, the wavelength-dependent permittivity of the shell was set to the one of silver sulfide.\cite{PetterssonTSF95}

For particle \#20 we scaled down the mesh by a factor of $0.97$, creating a layer of thickness $h=1.0$\,nm. For particle \#3 we used a scaling factor of $0.985$, yielding $h=0.3$\,nm. These shell thicknesses yield a good agreement between simulated and measured scattering cross-section spectra, as shown in \autoref{fig:fig6}.. For the absorption cross-section spectra, which are increased by the tarnish, some mismatch remains. \add{We show in the ESI \autoref{S-sec:tarnish_composition} that assuming silver oxide instead of silver sulfide, a similar effect on the cross-sections is found for a slightly larger thickness. Importantly,} we note that a tarnish layer can have a much more complex morphology than assumed here, and can also contain a mixture of sulfide, oxide, and even other compounds such as FeS. A residual mismatch is therefore expected considering the simple tarnish model employed. \add{We emphasize that for some of the NPs (e.g. \#12, see ESI \autoref{S-fig:particles7to12}), there is a good agreement between measured and simulated spectra within the expected uncertainty from the shape reconstruction (see ESI \autoref{S-fig:geometry}), without adjustable parameters, indicating that the formation of a tarnish layer varies between particles even within the same preparation and TEM grid.}

\section{Conclusions}
\label{sec:conclusions}
We have used the pipeline for correlative and quantitative optical and structural electron microscopy characterization that we have recently developed \cite{WangNSA20} to study individual silver nanohedra synthesized by photochemistry. Importantly, we extended the method to include electron tomography, to determine the volumetric shape of the particles accurately, and used the resulting morphology and orientation for simulations of the quantitative optical cross-section spectra, for a fitting-parameter free quantitative comparison with the measured spectra. This is the first study of this type, combining fully quantitative optical cross-section measurements with correlative electron tomography determining quantitatively the 3D particle morphology and orientation, and corresponding quantitative simulations.

While generally a good agreement of simulated and measured cross-sections is found, quantitative differences are revealed. Specifically a red shift of the measurements compared to the simulations by a few percent, mostly for the larger particles, and a difference in magnitude, mostly a reduction for the smaller particles.
To understand the origin of the deviations, the influence of three aspects was investigated.
(i) The tomographic reconstruction method was examined, showing resulting morphology variations mostly for the smallest particles investigated. 
(ii) The addition of a realistic surface damping in the permittivity resulted in only slightly modified spectra.
(iii) Adding a thin surface layer of tarnish, here modelled as silver sulfide, brought about, for realistic thicknesses in the 1\,nm range, a match within the expected systematic errors. Let us emphasize that such conclusions would have been less stringent without the information on the cross-section magnitude. For instance, the red shift of the measured spectra can be explained both in terms of the geometry being more sharp, within the reconstruction accuracy, and by the tarnish layer; but only the latter hypothesis is in agreement with the measured cross-section magnitudes. 

The accuracy of the method can be improved going forward. For example, one could add polarization-dependent measurements and simulations, using linearly, radially, and azimuthally polarised light, where the latter has the advantage of only in-plane polarized excitation for both BF and DF, thus exciting the same resonances. Furthermore, the slight angular dependence of the objective transmission could be calibrated and taken into account. To avoid the formation of a tarnish layer, a similar study on gold nanohedra could be envisaged, allowing to isolate the accuracy of geometry and permittivity. 

This work and the adoption of the developed methodology paves the way towards an accurate quantitative understanding and verification of the morphology - optical response relation in plasmonic nanoparticles,
especially for particles with complex shapes, which are important building blocks for next-generation devices.

\section*{Author contribution}
Y.W., P.B., and W.L. developed the workflow for correlative electron and optical microscopy.
A.Z., P.B., and W.L. developed the quantitative optical microspectroscopy technique.
Y.W. performed the sample preparation and optical microspectroscopy.
A.Z., Z.S., P.B., and W.L. developed the numerical model and methods.
Z.S. and A.Z. performed the numerical simulations.
W.A. and S.B. performed electron tomography and analysis.
A.Z., Z.S., and W.L. wrote the initial draft.
All authors took part in interpreting the data and writing the manuscript.

CRediT: Conceptualization W.L.,P.B.,W.A.; Data curation Y.W.,Z.S.,A.Z.,W.A.,W.L.; Formal analysis Y.W., Z.S., A.Z., W.A., W.L.; Funding acquisition W.A., S.B., P.B., W.L.; Investigation Y.W., Z.S.,W.A.; Methodology Y.W., Z.S., A.Z., W.A.,P.B.,W.L.; Project administration W.A., S.B., P.B., W.L.; Resources S.B., P.B., W.L.; Software Z.S., A.Z.,W.L.; Supervision A.Z.,S.B.,P.B.,W.L.; Validation Y.W., Z.S., A.Z., W.A.; Visualization Y.W.,Z.S.,A.Z.,W.A.,W.L.; Writing – original draft Y.W.,Z.S.,A.Z.,W.A.,W.L.; Writing – review \& editing Y.W., Z.S., A.Z., W.A., S.B., P.B., W.L.\,.
\section*{Data availability}
Information about the data created during this research, including
how to access it, is available on the Cardiff University data archive
at \href{http://doi.org/10.17035/d.2022.0176556287}{http://doi.org/10.17035/d.2022.0176556287}.
\section*{Conflicts of interest}
The authors declare no conflicts of interest.
\section*{Acknowledgements}
Z.S. acknowledges the UK Engineering and Physical Sciences Research Council (EPSRC) for his Ph.D. studentship award (grant EP/R513003/1). Y.W. acknowledges Iwan Moreels (University of Ghent) for training in nanoparticle synthesis. Y. W. acknowledges the Biotechnology and Biological Sciences Research Council (BBSRC) for his Ph.D. studentship award (grant BB/L015889/1). This work was supported by the UK EPSRC (grants EP/I005072/1 and EP/M028313/1), and by the European Commission (EUSMI E191000350). W.A. acknowledges an Individual Fellowship from the Marie Sk{\l}odowska-Curie actions (MSCA) under the EU’s Horizon 2020 program (Grant 797153, SOPMEN). We thank Lukas Payne and Iestyn Pope for contributions to the development of the hardware and software used for the optical measurements.

%
%
\balance%
%
%

\begin{mcitethebibliography}{40}
	\providecommand*{\natexlab}[1]{#1}
	\providecommand*{\mciteSetBstSublistMode}[1]{}
	\providecommand*{\mciteSetBstMaxWidthForm}[2]{}
	\providecommand*{\mciteBstWouldAddEndPuncttrue}
	{\def\EndOfBibitem{\unskip.}}
	\providecommand*{\mciteBstWouldAddEndPunctfalse}
	{\let\EndOfBibitem\relax}
	\providecommand*{\mciteSetBstMidEndSepPunct}[3]{}
	\providecommand*{\mciteSetBstSublistLabelBeginEnd}[3]{}
	\providecommand*{\EndOfBibitem}{}
	\mciteSetBstSublistMode{f}
	\mciteSetBstMaxWidthForm{subitem}
	{(\emph{\alph{mcitesubitemcount}})}
	\mciteSetBstSublistLabelBeginEnd{\mcitemaxwidthsubitemform\space}
	{\relax}{\relax}
	
	\bibitem[Trügler(2016)]{TruglerBook16}
	A.~Trügler, \emph{Optical properties of metallic nanoparticles : basic
		principles and simulation}, Springer, Cham, 2016\relax
	\mciteBstWouldAddEndPuncttrue
	\mciteSetBstMidEndSepPunct{\mcitedefaultmidpunct}
	{\mcitedefaultendpunct}{\mcitedefaultseppunct}\relax
	\EndOfBibitem
	\bibitem[Lee \emph{et~al.}(2020)Lee, Kim, Im, Balamurugan, and Nam]{LeeNS20}
	Y.~Y. Lee, R.~M. Kim, S.~W. Im, M.~Balamurugan and K.~T. Nam, \emph{Nanoscale},
	2020, \textbf{12}, 58--66\relax
	\mciteBstWouldAddEndPuncttrue
	\mciteSetBstMidEndSepPunct{\mcitedefaultmidpunct}
	{\mcitedefaultendpunct}{\mcitedefaultseppunct}\relax
	\EndOfBibitem
	\bibitem[Bohren and Huffman(1998)]{BohrenBook98}
	C.~F. Bohren and D.~R. Huffman, \emph{Absorption and Scattering of Light by
		Small Particles}, Wiley-VCH Verlag, 1998\relax
	\mciteBstWouldAddEndPuncttrue
	\mciteSetBstMidEndSepPunct{\mcitedefaultmidpunct}
	{\mcitedefaultendpunct}{\mcitedefaultseppunct}\relax
	\EndOfBibitem
	\bibitem[Crut \emph{et~al.}(2014)Crut, Maioli, {Del Fatti}, and
	Vall{\'{e}}e]{CrutCSR14}
	A.~Crut, P.~Maioli, N.~{Del Fatti} and F.~Vall{\'{e}}e, \emph{Chem. Soc. Rev.},
	2014, \textbf{43}, 3921--3956\relax
	\mciteBstWouldAddEndPuncttrue
	\mciteSetBstMidEndSepPunct{\mcitedefaultmidpunct}
	{\mcitedefaultendpunct}{\mcitedefaultseppunct}\relax
	\EndOfBibitem
	\bibitem[Olson \emph{et~al.}(2015)Olson, Dominguez-Medina, Hoggard, Wang,
	Chang, and Link]{OlsonCSR15}
	J.~Olson, S.~Dominguez-Medina, A.~Hoggard, L.-Y. Wang, W.-S. Chang and S.~Link,
	\emph{Chem. Soc. Rev.}, 2015, \textbf{44}, 40--57\relax
	\mciteBstWouldAddEndPuncttrue
	\mciteSetBstMidEndSepPunct{\mcitedefaultmidpunct}
	{\mcitedefaultendpunct}{\mcitedefaultseppunct}\relax
	\EndOfBibitem
	\bibitem[Husnik \emph{et~al.}(2012)Husnik, Linden, Diehl, Niegemann, Busch, and
	Wegener]{HusnikPRL12}
	M.~Husnik, S.~Linden, R.~Diehl, J.~Niegemann, K.~Busch and M.~Wegener,
	\emph{Phys. Rev. Lett.}, 2012, \textbf{109}, 233902\relax
	\mciteBstWouldAddEndPuncttrue
	\mciteSetBstMidEndSepPunct{\mcitedefaultmidpunct}
	{\mcitedefaultendpunct}{\mcitedefaultseppunct}\relax
	\EndOfBibitem
	\bibitem[Khadir \emph{et~al.}(2020)Khadir, Andr{\'{e}}n, Chaumet, Monneret,
	Bonod, Käll, Sentenac, and Baffou]{KhadirO20}
	S.~Khadir, D.~Andr{\'{e}}n, P.~C. Chaumet, S.~Monneret, N.~Bonod, M.~Käll,
	A.~Sentenac and G.~Baffou, \emph{Optica}, 2020, \textbf{7}, 243--248\relax
	\mciteBstWouldAddEndPuncttrue
	\mciteSetBstMidEndSepPunct{\mcitedefaultmidpunct}
	{\mcitedefaultendpunct}{\mcitedefaultseppunct}\relax
	\EndOfBibitem
	\bibitem[Henry \emph{et~al.}(2011)Henry, Bingham, Ringe, Marks, Schatz, and
	Duyne]{HenryJPCC11}
	A.-I. Henry, J.~M. Bingham, E.~Ringe, L.~D. Marks, G.~C. Schatz and R.~P.~V.
	Duyne, \emph{J.~Phys. Chem.~C}, 2011, \textbf{115}, 9291--9305\relax
	\mciteBstWouldAddEndPuncttrue
	\mciteSetBstMidEndSepPunct{\mcitedefaultmidpunct}
	{\mcitedefaultendpunct}{\mcitedefaultseppunct}\relax
	\EndOfBibitem
	\bibitem[Perassi \emph{et~al.}(2010)Perassi, Hernandez-Garrido, Moreno, Encina,
	Coronado, and Midgley]{PerassiNL10}
	E.~M. Perassi, J.~C. Hernandez-Garrido, M.~S. Moreno, E.~R. Encina, E.~A.
	Coronado and P.~A. Midgley, \emph{Nano Lett.}, 2010, \textbf{10},
	2097--2104\relax
	\mciteBstWouldAddEndPuncttrue
	\mciteSetBstMidEndSepPunct{\mcitedefaultmidpunct}
	{\mcitedefaultendpunct}{\mcitedefaultseppunct}\relax
	\EndOfBibitem
	\bibitem[Perassi \emph{et~al.}(2014)Perassi, Hrelescu, Wisnet, Döblinger,
	Scheu, Jäckel, Coronado, and Feldmann]{PerassiACSN14}
	E.~M. Perassi, C.~Hrelescu, A.~Wisnet, M.~Döblinger, C.~Scheu, F.~Jäckel,
	E.~A. Coronado and J.~Feldmann, \emph{{ACS} Nano}, 2014, \textbf{8},
	4395--4402\relax
	\mciteBstWouldAddEndPuncttrue
	\mciteSetBstMidEndSepPunct{\mcitedefaultmidpunct}
	{\mcitedefaultendpunct}{\mcitedefaultseppunct}\relax
	\EndOfBibitem
	\bibitem[Hern{\'{a}}ndez-Garrido \emph{et~al.}(2014)Hern{\'{a}}ndez-Garrido,
	Moreno, Ducati, P{\'{e}}rez, Midgley, and Coronado]{HernandezNS14}
	J.~C. Hern{\'{a}}ndez-Garrido, M.~S. Moreno, C.~Ducati, L.~A. P{\'{e}}rez,
	P.~A. Midgley and E.~A. Coronado, \emph{Nanoscale}, 2014, \textbf{6},
	12696--12702\relax
	\mciteBstWouldAddEndPuncttrue
	\mciteSetBstMidEndSepPunct{\mcitedefaultmidpunct}
	{\mcitedefaultendpunct}{\mcitedefaultseppunct}\relax
	\EndOfBibitem
	\bibitem[Chuntonov \emph{et~al.}(2011)Chuntonov, Bar-Sadan, Houben, and
	Haran]{ChuntonovNL11}
	L.~Chuntonov, M.~Bar-Sadan, L.~Houben and G.~Haran, \emph{Nano Lett.}, 2011,
	\textbf{12}, 145--150\relax
	\mciteBstWouldAddEndPuncttrue
	\mciteSetBstMidEndSepPunct{\mcitedefaultmidpunct}
	{\mcitedefaultendpunct}{\mcitedefaultseppunct}\relax
	\EndOfBibitem
	\bibitem[Zilli \emph{et~al.}(2019)Zilli, Langbein, and Borri]{ZilliACSP19}
	A.~Zilli, W.~Langbein and P.~Borri, \emph{{ACS} Photonics}, 2019, \textbf{6},
	2149--2160\relax
	\mciteBstWouldAddEndPuncttrue
	\mciteSetBstMidEndSepPunct{\mcitedefaultmidpunct}
	{\mcitedefaultendpunct}{\mcitedefaultseppunct}\relax
	\EndOfBibitem
	\bibitem[Borri \emph{et~al.}(2017)Borri, Langbein, Zilli, and
	Payne]{BorriUSpatent21}
	P.~Borri, W.~W. Langbein, A.~Zilli and L.~M. Payne, \emph{Analysing
		nano-objects}, 2017, U.S. patent 10996159\relax
	\mciteBstWouldAddEndPuncttrue
	\mciteSetBstMidEndSepPunct{\mcitedefaultmidpunct}
	{\mcitedefaultendpunct}{\mcitedefaultseppunct}\relax
	\EndOfBibitem
	\bibitem[Wang \emph{et~al.}(2020)Wang, Zilli, Sztranyovszky, Langbein, and
	Borri]{WangNSA20}
	Y.~Wang, A.~Zilli, Z.~Sztranyovszky, W.~Langbein and P.~Borri, \emph{Nanoscale
		Adv.}, 2020, \textbf{2}, 2485--2496\relax
	\mciteBstWouldAddEndPuncttrue
	\mciteSetBstMidEndSepPunct{\mcitedefaultmidpunct}
	{\mcitedefaultendpunct}{\mcitedefaultseppunct}\relax
	\EndOfBibitem
	\bibitem[Pietrobon and Kitaev(2008)]{PietrobonCM08}
	B.~Pietrobon and V.~Kitaev, \emph{Chem. Mater.}, 2008, \textbf{20},
	5186--5190\relax
	\mciteBstWouldAddEndPuncttrue
	\mciteSetBstMidEndSepPunct{\mcitedefaultmidpunct}
	{\mcitedefaultendpunct}{\mcitedefaultseppunct}\relax
	\EndOfBibitem
	\bibitem[Zheng \emph{et~al.}(2009)Zheng, Zhao, Guo, Tang, Xu, Zhao, Xu, and
	Lombardi]{ZhengL09}
	X.~Zheng, X.~Zhao, D.~Guo, B.~Tang, S.~Xu, B.~Zhao, W.~Xu and J.~R. Lombardi,
	\emph{Langmuir}, 2009, \textbf{25}, 3802--3807\relax
	\mciteBstWouldAddEndPuncttrue
	\mciteSetBstMidEndSepPunct{\mcitedefaultmidpunct}
	{\mcitedefaultendpunct}{\mcitedefaultseppunct}\relax
	\EndOfBibitem
	\bibitem[Ye \emph{et~al.}(2015)Ye, Song, Tian, Chen, Wang, Niu, and Qu]{YeNS15}
	S.~Ye, J.~Song, Y.~Tian, L.~Chen, D.~Wang, H.~Niu and J.~Qu, \emph{Nanoscale},
	2015, \textbf{7}, 12706--12712\relax
	\mciteBstWouldAddEndPuncttrue
	\mciteSetBstMidEndSepPunct{\mcitedefaultmidpunct}
	{\mcitedefaultendpunct}{\mcitedefaultseppunct}\relax
	\EndOfBibitem
	\bibitem[Das and Chini(2012)]{DasJPCC12}
	P.~Das and T.~K. Chini, \emph{J.~Phys. Chem.~C}, 2012, \textbf{116},
	25969--25976\relax
	\mciteBstWouldAddEndPuncttrue
	\mciteSetBstMidEndSepPunct{\mcitedefaultmidpunct}
	{\mcitedefaultendpunct}{\mcitedefaultseppunct}\relax
	\EndOfBibitem
	\bibitem[Myroshnychenko \emph{et~al.}(2012)Myroshnychenko, Nelayah, Adamo,
	Geuquet, Rodr{\'{\i}}guez-Fern{\'{a}}ndez, Pastoriza-Santos, MacDonald,
	Henrard, Liz-Marz{\'{a}}n, Zheludev, Kociak, and {Garc{\'{\i}}a de
		Abajo}]{MyroshnychenkoNL12}
	V.~Myroshnychenko, J.~Nelayah, G.~Adamo, N.~Geuquet,
	J.~Rodr{\'{\i}}guez-Fern{\'{a}}ndez, I.~Pastoriza-Santos, K.~F. MacDonald,
	L.~Henrard, L.~M. Liz-Marz{\'{a}}n, N.~I. Zheludev, M.~Kociak and F.~J.
	{Garc{\'{\i}}a de Abajo}, \emph{Nano Lett.}, 2012, \textbf{12},
	4172--4180\relax
	\mciteBstWouldAddEndPuncttrue
	\mciteSetBstMidEndSepPunct{\mcitedefaultmidpunct}
	{\mcitedefaultendpunct}{\mcitedefaultseppunct}\relax
	\EndOfBibitem
	\bibitem[Alekseeva \emph{et~al.}(2019)Alekseeva, Nedrygailov, and
	Langhammer]{AlekseevaACSP19}
	S.~Alekseeva, I.~I. Nedrygailov and C.~Langhammer, \emph{{ACS} Photonics},
	2019, \textbf{6}, 1319--1330\relax
	\mciteBstWouldAddEndPuncttrue
	\mciteSetBstMidEndSepPunct{\mcitedefaultmidpunct}
	{\mcitedefaultendpunct}{\mcitedefaultseppunct}\relax
	\EndOfBibitem
	\bibitem[Liao \emph{et~al.}(2019)Liao, Fang, Li, Li, Xu, and Fang]{LiaoNS19}
	G.~Liao, J.~Fang, Q.~Li, S.~Li, Z.~Xu and B.~Fang, \emph{Nanoscale}, 2019,
	\textbf{11}, 7062--7096\relax
	\mciteBstWouldAddEndPuncttrue
	\mciteSetBstMidEndSepPunct{\mcitedefaultmidpunct}
	{\mcitedefaultendpunct}{\mcitedefaultseppunct}\relax
	\EndOfBibitem
	\bibitem[Kang \emph{et~al.}(2018)Kang, Buchman, Rodriguez, Ring, He, Bantz, and
	Haynes]{KangCR18}
	H.~Kang, J.~T. Buchman, R.~S. Rodriguez, H.~L. Ring, J.~He, K.~C. Bantz and
	C.~L. Haynes, \emph{Chem. Rev.}, 2018, \textbf{119}, 664--699\relax
	\mciteBstWouldAddEndPuncttrue
	\mciteSetBstMidEndSepPunct{\mcitedefaultmidpunct}
	{\mcitedefaultendpunct}{\mcitedefaultseppunct}\relax
	\EndOfBibitem
	\bibitem[Payne \emph{et~al.}(2019)Payne, Zilli, Wang, Langbein, and
	Borri]{PayneSPIE19}
	L.~Payne, A.~Zilli, Y.~Wang, W.~Langbein and P.~Borri, Proc. SPIE 10892,
	Colloidal Nanoparticles for Biomedical Applications {XIV}, 2019\relax
	\mciteBstWouldAddEndPuncttrue
	\mciteSetBstMidEndSepPunct{\mcitedefaultmidpunct}
	{\mcitedefaultendpunct}{\mcitedefaultseppunct}\relax
	\EndOfBibitem
	\bibitem[Moon(1996)]{MoonSPM96}
	T.~K. Moon, \emph{{IEEE} Signal Process. Mag.}, 1996, \textbf{13}, 47--60\relax
	\mciteBstWouldAddEndPuncttrue
	\mciteSetBstMidEndSepPunct{\mcitedefaultmidpunct}
	{\mcitedefaultendpunct}{\mcitedefaultseppunct}\relax
	\EndOfBibitem
	\bibitem[van Aarle \emph{et~al.}(2015)van Aarle, Palenstijn, Beenhouwer,
	Altantzis, Bals, Batenburg, and Sijbers]{AarleU15}
	W.~van Aarle, W.~J. Palenstijn, J.~D. Beenhouwer, T.~Altantzis, S.~Bals, K.~J.
	Batenburg and J.~Sijbers, \emph{Ultramicroscopy}, 2015, \textbf{157},
	35--47\relax
	\mciteBstWouldAddEndPuncttrue
	\mciteSetBstMidEndSepPunct{\mcitedefaultmidpunct}
	{\mcitedefaultendpunct}{\mcitedefaultseppunct}\relax
	\EndOfBibitem
	\bibitem[Yang \emph{et~al.}(2015)Yang, D{\textquotesingle}Archangel,
	Sundheimer, Tucker, Boreman, and Raschke]{YangPRB15}
	H.~U. Yang, J.~D{\textquotesingle}Archangel, M.~L. Sundheimer, E.~Tucker, G.~D.
	Boreman and M.~B. Raschke, \emph{Phys. Rev.~B}, 2015, \textbf{91},
	235137\relax
	\mciteBstWouldAddEndPuncttrue
	\mciteSetBstMidEndSepPunct{\mcitedefaultmidpunct}
	{\mcitedefaultendpunct}{\mcitedefaultseppunct}\relax
	\EndOfBibitem
	\bibitem[Zhang \emph{et~al.}(2009)Zhang, Li, Wu, Schatz, and
	Mirkin]{ZhangACIE09}
	J.~Zhang, S.~Li, J.~Wu, G.~Schatz and C.~Mirkin, \emph{Angew. Chem. - Int.
		Ed.}, 2009, \textbf{48}, 7787--7791\relax
	\mciteBstWouldAddEndPuncttrue
	\mciteSetBstMidEndSepPunct{\mcitedefaultmidpunct}
	{\mcitedefaultendpunct}{\mcitedefaultseppunct}\relax
	\EndOfBibitem
	\bibitem[Grześkiewicz \emph{et~al.}(2014)Grześkiewicz, Ptaszyński, and
	Kotkowiak]{GrzeskiewiczP14}
	B.~Grześkiewicz, K.~Ptaszyński and M.~Kotkowiak, \emph{Plasmonics}, 2014,
	\textbf{9}, 607–614\relax
	\mciteBstWouldAddEndPuncttrue
	\mciteSetBstMidEndSepPunct{\mcitedefaultmidpunct}
	{\mcitedefaultendpunct}{\mcitedefaultseppunct}\relax
	\EndOfBibitem
	\bibitem[Raziman and Martin(2013)]{RazimanOE13}
	T.~V. Raziman and O.~J.~F. Martin, \emph{Opt. Express}, 2013, \textbf{21},
	21500–21507\relax
	\mciteBstWouldAddEndPuncttrue
	\mciteSetBstMidEndSepPunct{\mcitedefaultmidpunct}
	{\mcitedefaultendpunct}{\mcitedefaultseppunct}\relax
	\EndOfBibitem
	\bibitem[Rodríguez-Fernández \emph{et~al.}(2009)Rodríguez-Fernández, Novo,
	Myroshnychenko, Funston, Sánchez-Iglesias, Pastoriza-Santos, Pérez-Juste,
	García~de Abajo, Liz-Marzán, and Mulvaney]{RodriguezFernandezJPCC09}
	J.~Rodríguez-Fernández, C.~Novo, V.~Myroshnychenko, A.~M. Funston,
	A.~Sánchez-Iglesias, I.~Pastoriza-Santos, J.~Pérez-Juste, F.~J. García~de
	Abajo, L.~M. Liz-Marzán and P.~Mulvaney, \emph{J. Phys. Chem. C}, 2009,
	\textbf{113}, 18623–18631\relax
	\mciteBstWouldAddEndPuncttrue
	\mciteSetBstMidEndSepPunct{\mcitedefaultmidpunct}
	{\mcitedefaultendpunct}{\mcitedefaultseppunct}\relax
	\EndOfBibitem
	\bibitem[Marks and Peng(2016)]{MarksJPCM16}
	L.~D. Marks and L.~Peng, \emph{J. Phys.: Condens. Matter}, 2016, \textbf{28},
	053001\relax
	\mciteBstWouldAddEndPuncttrue
	\mciteSetBstMidEndSepPunct{\mcitedefaultmidpunct}
	{\mcitedefaultendpunct}{\mcitedefaultseppunct}\relax
	\EndOfBibitem
	\bibitem[Voisin \emph{et~al.}(2001)Voisin, Del~Fatti, Christofilos, and
	Vallée]{VoisinJPCB01}
	C.~Voisin, N.~Del~Fatti, D.~Christofilos and F.~Vallée, \emph{J. Phys. Chem.
		B}, 2001, \textbf{105}, 2264–2280\relax
	\mciteBstWouldAddEndPuncttrue
	\mciteSetBstMidEndSepPunct{\mcitedefaultmidpunct}
	{\mcitedefaultendpunct}{\mcitedefaultseppunct}\relax
	\EndOfBibitem
	\bibitem[Berciaud \emph{et~al.}(2005)Berciaud, Cognet, Tamarat, and
	Lounis]{BerciaudNL05}
	S.~Berciaud, L.~Cognet, P.~Tamarat and B.~Lounis, \emph{Nano Lett.}, 2005,
	\textbf{5}, 515–518\relax
	\mciteBstWouldAddEndPuncttrue
	\mciteSetBstMidEndSepPunct{\mcitedefaultmidpunct}
	{\mcitedefaultendpunct}{\mcitedefaultseppunct}\relax
	\EndOfBibitem
	\bibitem[Masia \emph{et~al.}(2012)Masia, Langbein, and Borri]{MasiaPRB12}
	F.~Masia, W.~Langbein and P.~Borri, \emph{Phys. Rev. B}, 2012, \textbf{85},
	235403\relax
	\mciteBstWouldAddEndPuncttrue
	\mciteSetBstMidEndSepPunct{\mcitedefaultmidpunct}
	{\mcitedefaultendpunct}{\mcitedefaultseppunct}\relax
	\EndOfBibitem
	\bibitem[Muskens \emph{et~al.}(2008)Muskens, Billaud, Broyer, Del~Fatti, and
	Vallée]{MuskensPRB08}
	O.~L. Muskens, P.~Billaud, M.~Broyer, N.~Del~Fatti and F.~Vallée, \emph{Phys.
		Rev. B}, 2008, \textbf{78}, 205410\relax
	\mciteBstWouldAddEndPuncttrue
	\mciteSetBstMidEndSepPunct{\mcitedefaultmidpunct}
	{\mcitedefaultendpunct}{\mcitedefaultseppunct}\relax
	\EndOfBibitem
	\bibitem[Mcmahon \emph{et~al.}(2005)Mcmahon, Lopez, Meyer, Feldman, and
	Haglund]{McmahonAPB05}
	M.~D. Mcmahon, R.~Lopez, H.~M. Meyer, L.~C. Feldman and R.~F. Haglund,
	\emph{Appl. Phys.~B}, 2005, \textbf{80}, 915--921\relax
	\mciteBstWouldAddEndPuncttrue
	\mciteSetBstMidEndSepPunct{\mcitedefaultmidpunct}
	{\mcitedefaultendpunct}{\mcitedefaultseppunct}\relax
	\EndOfBibitem
	\bibitem[Elechiguerra \emph{et~al.}(2005)Elechiguerra, Larios-Lopez, Liu,
	Garcia-Gutierrez, Camacho-Bragado, and Yacaman]{ElechiguerraCM05}
	J.~L. Elechiguerra, L.~Larios-Lopez, C.~Liu, D.~Garcia-Gutierrez,
	A.~Camacho-Bragado and M.~J. Yacaman, \emph{Chem. Mater.}, 2005, \textbf{17},
	6042--6052\relax
	\mciteBstWouldAddEndPuncttrue
	\mciteSetBstMidEndSepPunct{\mcitedefaultmidpunct}
	{\mcitedefaultendpunct}{\mcitedefaultseppunct}\relax
	\EndOfBibitem
	\bibitem[Grillet \emph{et~al.}(2013)Grillet, Manchon, Cottancin, Bertorelle,
	Bonnet, Broyer, Lerm{\'{e}}, and Pellarin]{GrilletJPCC13}
	N.~Grillet, D.~Manchon, E.~Cottancin, F.~Bertorelle, C.~Bonnet, M.~Broyer,
	J.~Lerm{\'{e}} and M.~Pellarin, \emph{J. Phys. Chem. C}, 2013, \textbf{117},
	2274--2282\relax
	\mciteBstWouldAddEndPuncttrue
	\mciteSetBstMidEndSepPunct{\mcitedefaultmidpunct}
	{\mcitedefaultendpunct}{\mcitedefaultseppunct}\relax
	\EndOfBibitem
	\bibitem[Pettersson and Snyder(1995)]{PetterssonTSF95}
	L.~Pettersson and P.~Snyder, \emph{Thin Solid Films}, 1995, \textbf{270},
	69--72\relax
	\mciteBstWouldAddEndPuncttrue
	\mciteSetBstMidEndSepPunct{\mcitedefaultmidpunct}
	{\mcitedefaultendpunct}{\mcitedefaultseppunct}\relax
	\EndOfBibitem
\end{mcitethebibliography}
%

\providecommand*{\mcitethebibliography}{\thebibliography}
\csname @ifundefined\endcsname{endmcitethebibliography}
{\let\endmcitethebibliography\endthebibliography}{}

\end{document}


%
\title{Electronic supplementary information to:
	\texorpdfstring{\\}{}
	Quantitatively linking morphology and optical response of individual silver nanohedra
	}
\date{\today}
\pacs{}			
\keywords{}		
\author{Yisu Wang}
	\thanks{These authors contributed in equal measure to this work.}
	\affiliation{School of Biosciences, Cardiff University -- Museum Avenue, Cardiff CF10 3AX, UK}
\author{Zoltan Sztranyovszky}
	\thanks{These authors contributed in equal measure to this work.}
	\affiliation{School of Physics and Astronomy, Cardiff University -- The Parade, Cardiff CF24 3AA, UK}
\author{Attilio Zilli}
\affiliation{School of Biosciences, Cardiff University -- Museum Avenue, Cardiff CF10 3AX, UK}
\affiliation{Department of Physics, Politecnico di Milano -- Piazza Leonardo da Vinci 32, 20133 Milano, Italy}
\author{Wiebke Albrecht}
\thanks{Present address:\\[-3pt] Department of Sustainable Energy Materials, AMOLF, Science Park 104, 1098 XG Amsterdam, The Netherlands.}
\affiliation{EMAT and NANOlab Center of Excellence, University of Antwerp -- Groenenborgerlaan 171, B-2020 Antwerp, Belgium}
\author{Sara Bals}
\affiliation{EMAT and NANOlab Center of Excellence, University of Antwerp -- Groenenborgerlaan 171, B-2020 Antwerp, Belgium}
\author{Paola Borri}
	\affiliation{School of Biosciences, Cardiff University -- Museum Avenue, Cardiff CF10 3AX, UK}
\author{Wolfgang Langbein}
	\email{langbeinww@cardiff.ac.uk}
	\affiliation{School of Physics and Astronomy, Cardiff University -- The Parade, Cardiff CF24 3AA, UK}
\maketitle 

\section{Sample fabrication\label{sec:sample_fabrication}}

\begin{figure}[b]
	\includegraphics[width=0.8\textwidth]{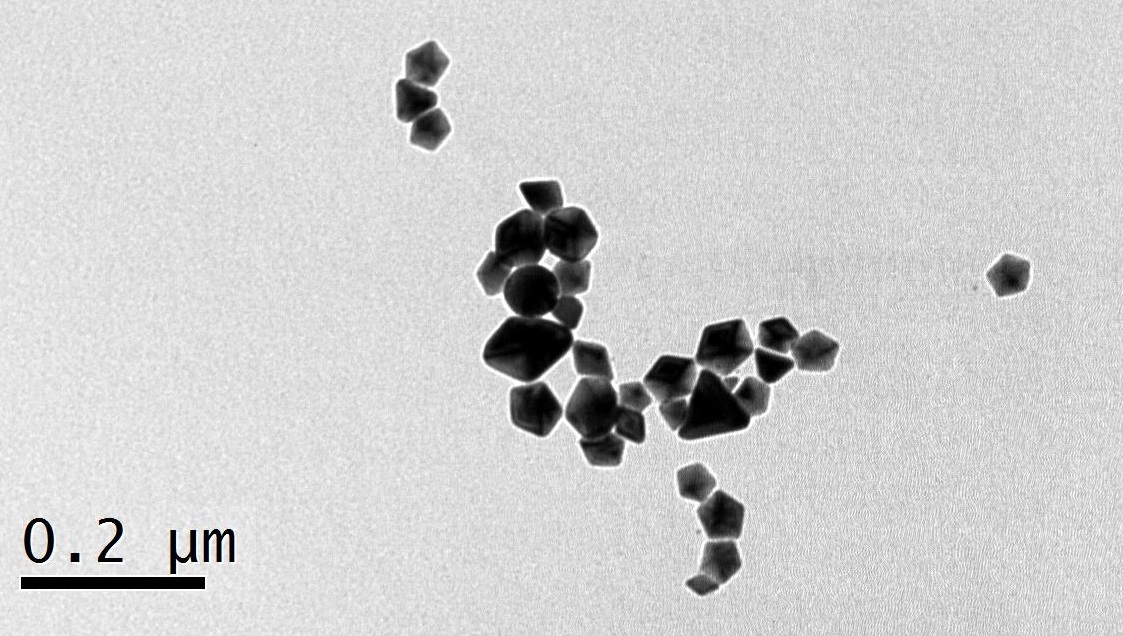}
	\caption{\label{fig:hrtem} HR-TEM images (JEOL JEM-1011) of the drop-casted purified product solution.}
\end{figure}

The fabrication procedure is based on \Onlinecite{PietrobonCM08}. The seeds were fabricated by reduction of silver nitrate (AgNO$_3$) in aqueous solution. 8\,mL seed solution was prepared by mixing 0.5\,\textmu M silver nitrate (Sigma Aldrich), 6.25 \textmu M polyvinylpyrrolidone (PVP) of molecular weight 10k (Sigma Aldrich), 3\,mM trisodium citrate (Na$_3$C$_6$H$_5$O$_7$) (Sigma Aldrich) and 0.65\,\textmu M sodium hydroborate (NaBH$_4$) (Sigma Aldrich) with vigorous stir for 3\,mins at room temperature until the color of the solution turned into light yellow. The seed solution, placed in a glass vial (Fisherbrand, type 1 class A borosilicate glass) and covered with a glass coverslip (Agar Scientific \#1.5) was then irradiated for 7 hours at room temperature via a royal-blue (447\,nm) LUXEON Rebel ES LED with a measured optical power of 710\,mW in a home-built chamber with a cylindrical inner volume of 53\,mm diameter and 95\,mm height, made of aluminium. The inner surface of the chamber was painted first with a white primer (Starglow Universal Primer, Glowtec, UK) then by a reflective varnish (Starglow Clear Reflective Paint, Glowtec, UK) to achieve a high diffuse reflectivity ($>95$\%) improving the intensity and homogeneity of the irradiation.
The product solution after irradiation had an orange color, and was then purified via a 2-step centrifugation to minimise the aggregation in the pellet: a first step at 500 relative centrifugal force (RCF) for 20\,min (to remove the small silver crystals) was followed by a second step at 1500\,RCF for 20\,min. After each step, the supernatant was removed and the pellet was resuspended in 0.1\% PVP with 2\,mM trisodium citrate solution. The product solution was stable (seen by a stable colour) in the fridge at 4$^\circ$C  for several months. All nanoparticles (NPs) analysed were synthesised no more than two days two before optical imaging. Conventional high resolution transmission electron microscopy (HRTEM) images of the purified solution were acquired on a JEOL JEM-1011 microscope equipped with a thermionic gun at 100 kV accelerating voltage. Samples were prepared by drop-casting NP suspensions onto carbon film-coated 200 mesh copper grids. \autoref{fig:hrtem} shows the fabricated NPs, which are dominantly decahedra, but also include triangular plates, bipyramids, and other shapes. 

\begin{figure}[b]
	\includegraphics[width=\textwidth]{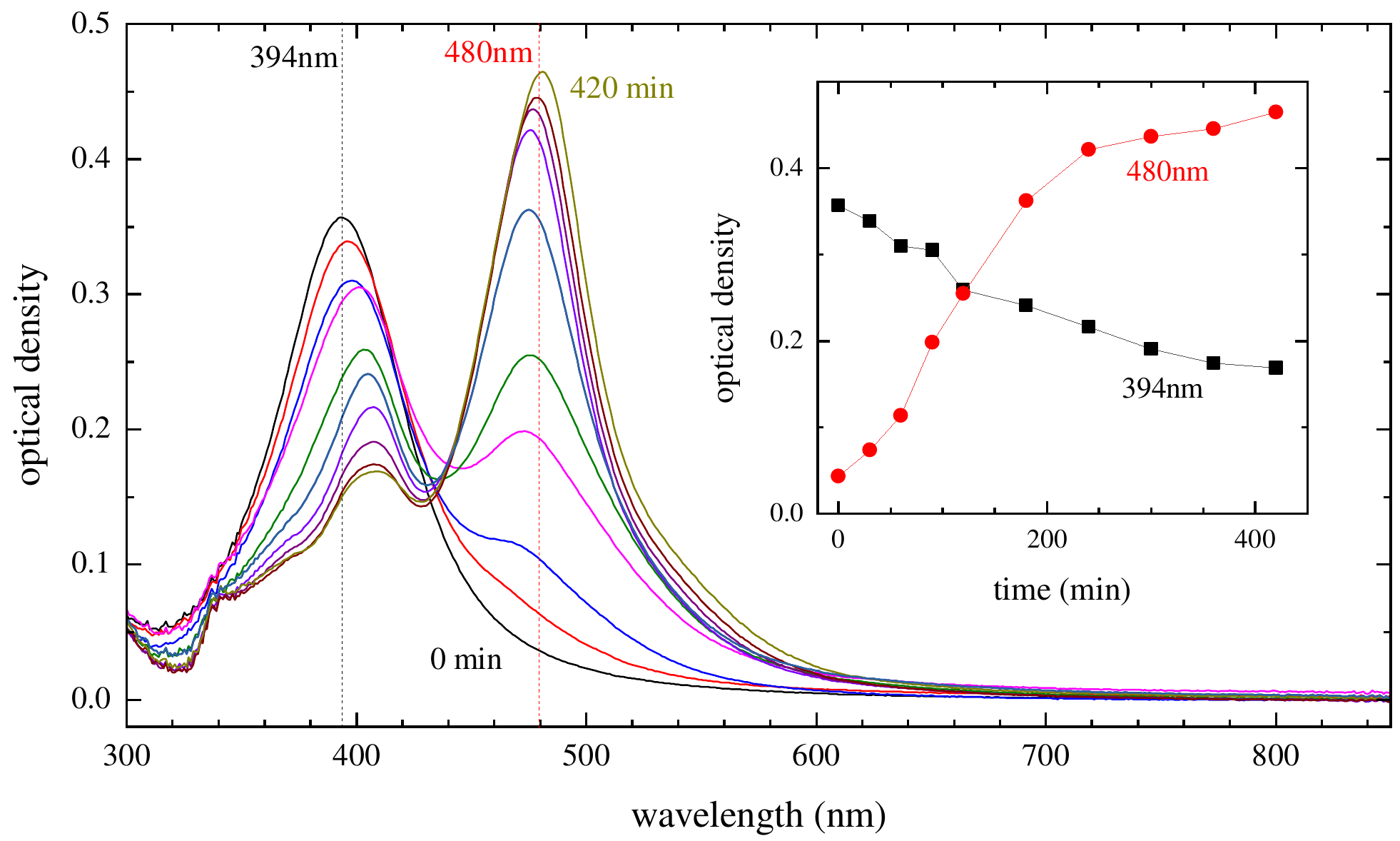}
	\caption{\label{fig:kinetic} Kinetic study of decahedra formation under 455\,nm LED irradiation: 7\,mL of seed solution (optical density 3.5 at 400\,nm) was irradiated with a LED (ThorLabs M455L3-C1) of 455\,nm wavelength and 500\,mW power with the sample and LED output in close proximity enclosed in aluminium foil. The irradiation was paused at different time points, and 50\,\textmu L of the reaction solution was sampled and diluted (1 in 10) for UV--Vis spectroscopy. Extinction spectra recorded using a Varian Cary 3000 UV--Vis spectrophotometer for various irradiation times are shown. The inset shows the measured optical density at 394\,nm and 490\,nm as a function of  irradiation time.}
\end{figure}

A preliminary kinetic study shown in \autoref{fig:kinetic} was performed to determine the formation times. 
The observed UV--Vis spectral evolution of the photochemical growth is consistent with data previously published  in \Onlinecite{ZhengL09}. A progressive rise of a plasmonic peak at 480\,nm is observed, which is characteristic of Ag decahedra in aqueous solution.

\clearpage

\section{Sulfidisation\label{sec:sulfidisation}}

In the main text we discuss the influence of a thin Ag$_2$S tarnish layer, leading to a better agreement between simulated and measured optical cross-sections. Such a layer can form due to exposure to trace amounts of sulfur, either in the surrounding atmosphere (H$_2$S for example), or on the TEM grid as residuals from the sample preparation.

In order to minimize contamination, the samples were shipped as follows: immediately after the optical measurements, the sample grids were placed in a standard TEM grid holder, and the holder was encapsulated in polypropylene centrifuge tubes filled with nitrogen, and rigorously sealed with parafilm. The tubes were shipped from Cardiff to Antwerp in room-temperature packaging via next-day delivery.
The samples were then measured within two days of arrival, and opened immediately before loading onto the electron microscope for imaging.

Notably, in a first round of experiments (not included in the results presented in the main text), the SiO$_2$ film of the TEM grid was cleaned and activated by the standard Piranha solution with reduced sulfuric acid (5\%). The grid was incubated for 1 hour at 55\,$^\circ$C in 10\,mL etching solution of 500\,\textmu l H$_2$SO$_4$ (99\%) diluted with 9.5\,mL of 30\% H$_2$O$_2$.

\begin{figure}[b]
	\includegraphics[width=\textwidth]{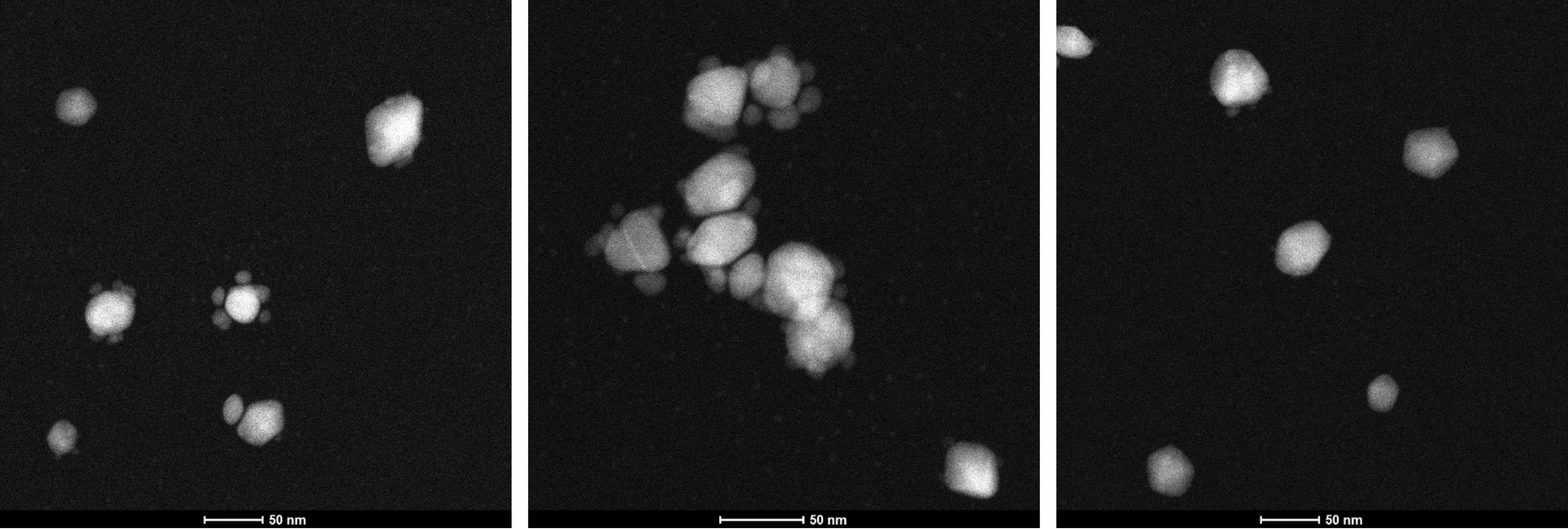}
	\caption{\label{fig:sulfidisation} High-angle annular dark-field scanning TEM (HAADF-STEM) images of Ag particles on the SiO$_2$ windows which were cleaned and activated with a protocol involving sulfuric acid in the first round of experiments, after optical characterisation. For most particles, a set of smaller surrounding debris is visible.}
\end{figure}

\begin{figure}
	\includegraphics[width=\textwidth]{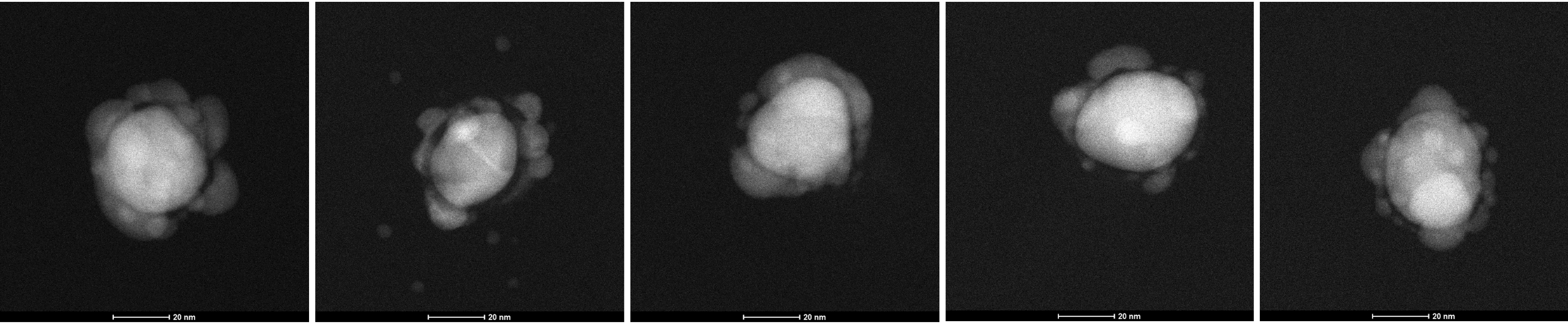}
	\caption{\label{fig:sulfidisation_p} Same as \autoref{fig:sulfidisation}, but zooms on selected Ag particles whose cross-section spectra showed a plasmon resonance peak.}
\end{figure}

\begin{figure}
	\includegraphics[width=\textwidth]{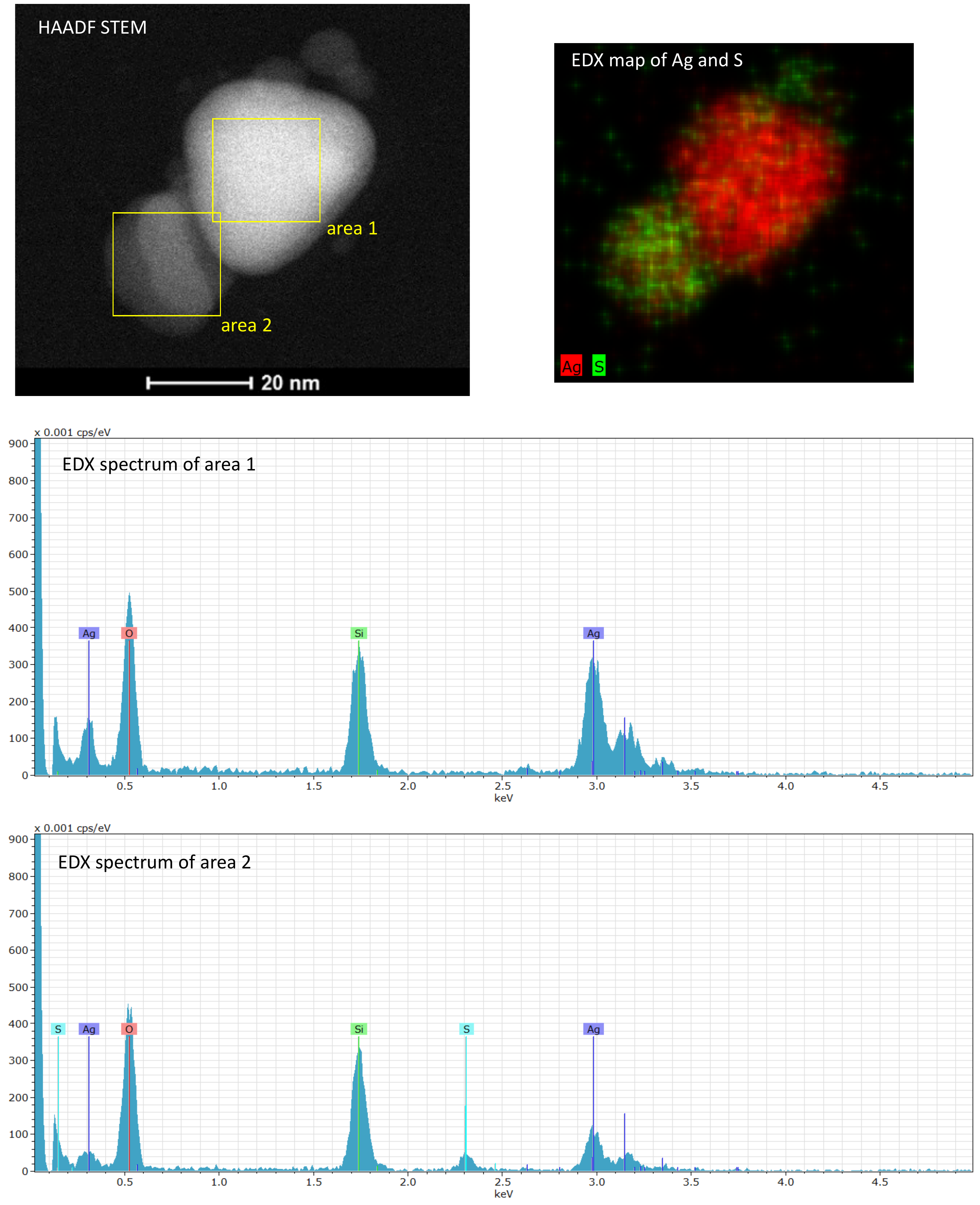}
	\caption{\label{fig:sulfidisation_EDX} EDX elemental analysis. Top left: HAADF-STEM image of Ag particles as in \autoref{fig:sulfidisation}. Two areas are indicated, whose EDX spectra are shown on the bottom. Top right: EDX map (smoothed with 3 pixels width) of the Ag peak net counts (red, $2.9-3.1\,$keV) and the S peak net counts (green, $2.2-2.4\,$keV).}
\end{figure}

While plasmonic NPs were identified in the optical measurements, subsequent electron microscopy indicated that many NPs in this batch	 were either completely converted to or were surrounded by debris containing sulfur (most likely Ag$_2$S) as shown in \autoref{fig:sulfidisation}. These also included NPs whose optical cross-section displayed plasmonic peaks in the preceding measurements(see \autoref{fig:sulfidisation_p}). Energy-dispersive X-ray (EDX) spectroscopy, shown in \autoref{fig:sulfidisation_EDX}, confirmed the sulfur content of the debris, revealed by a characteristic sulfur peak emerging at 2.5\,keV and the corresponding decrease of the Ag peaks at 0.3 and 3.0\,keV. EDX maps such as the one displayed in \autoref{fig:sulfidisation_EDX} top right were acquired using the Super-X detector of the Tecnai Osiris TEM operated at 200 kV. The maps were generally acquired for 10\,min at a current of 150\,pA. 

These findings in the first round of experiments indicated that the piranha solution might leave some H$_2$SO$_4$ on the grid, which might dissolve in anisole. After optical imaging the sample grid was held by a reverse-action tweezer and air-dried at 32$^\circ$C, which could allow the residual to deposit on the grid surface and the particles, promoting sulfidisation. 

\begin{figure}
	\includegraphics[width=\textwidth]{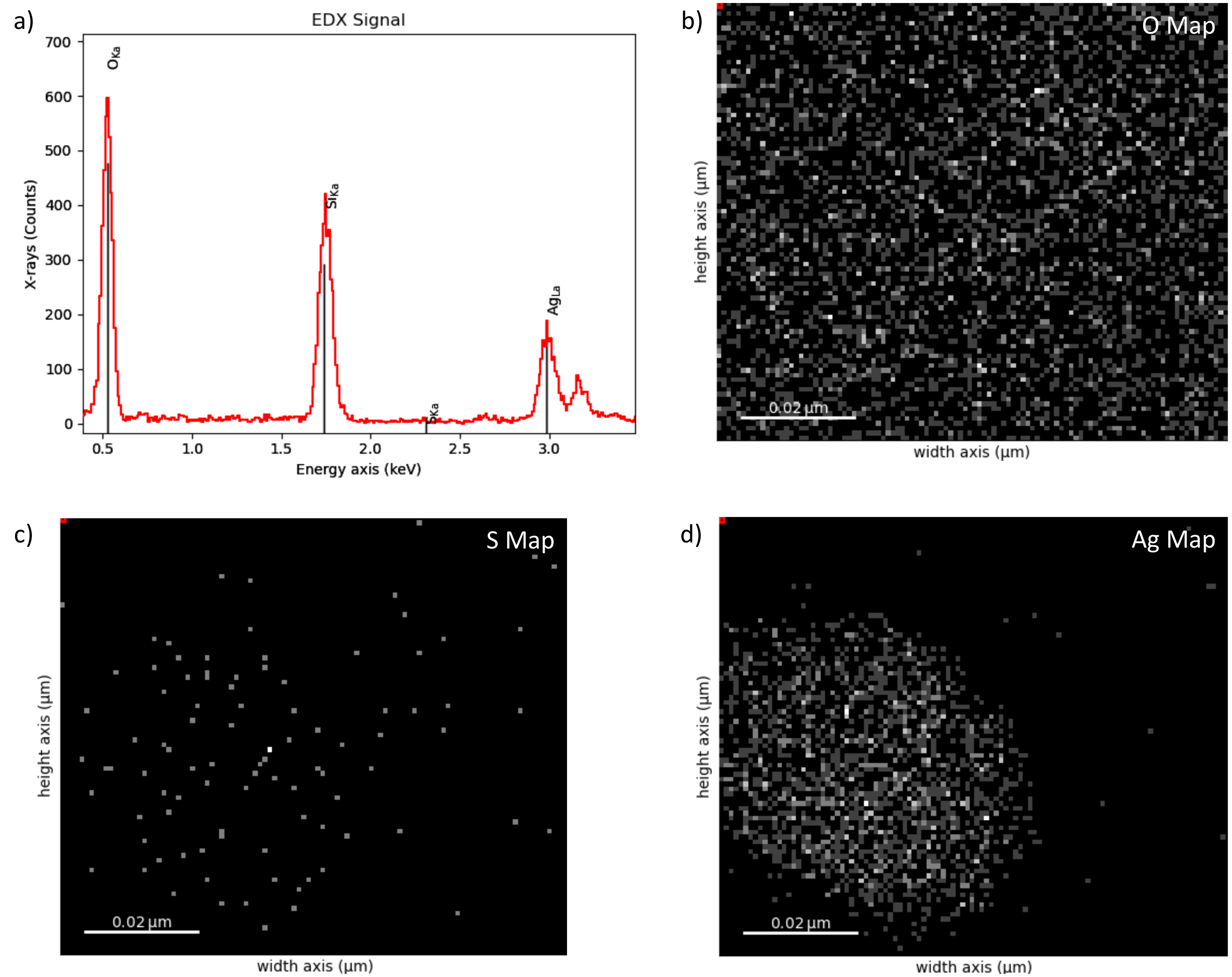}
	\caption{\label{fig:EDXBatch2} EDX elemental analysis of particle \#14. a) EDX spectrum. EDX maps of b) the Ag peak counts ($2.9-3.1\,$keV) c) the S peak counts ($2.2-2.4\,$keV), and d) the O peak counts ($0.4-0.6\,$keV).}
\end{figure}

In the second round of experiments, the sulfuric acid in the piranha solution was substituted with hydrochloric acid (see the \autoref{P-sec:experiment} of the main paper). We found that this change in the protocol still granted NP immobilization, but avoided the formation of the obvious debris around particles.  Since there was no structure observed in HAADF-STEM indicating a surface layer such as Ag$_2$S in the second round, EDX measurements were done only on a few particles. An example for particle \#14 is shown in \autoref{fig:EDXBatch2}. The summed spectrum of the whole area contains no clear S peak, and the S map shows only uncompensated background signal. These results show, that within the limited signal-to-noise ratio of EDX, no S could be detected. Compared to the first round of experiments (\hyperref[fig:sulfidisation]{Figs.\,\ref{fig:sulfidisation}} - \hyperref[fig:sulfidisation_EDX]{\ref{fig:sulfidisation_EDX}}), a possible Ag$_2$S surface layer must  be very thin. As the grid is made of silica, the oxygen signal does not allow to locate a possible Ag$_2$O layer. 

It should be noted that due to the low signal-to-noise ratio in EDX measurements, detecting thin (0.3--1.6\,nm) sulfide layers as introduced in the main text is a challenge. Achieving sufficient sensitivity in EDX maps requires long acquisition times at higher beam currents than HAADF-STEM imaging, which induces beam damage. As a result, the spatial distribution of the element maps would be questionable, not only due to particle reshaping but also due to the possibility that the chemical modification could have occurred during the EDX measurements by material released from the support.

Regarding the possible origin of the sulfur, we revisited the protocol used. A  thinkable origin could be the stainless steel reverse-action tweezers used to hold the grid during the cleaning and grid functionalisation process. They might contain small amounts of iron sulfide, which could react with the etchant used in a reaction FeS + 2 HCl → FeCl$_2$ + H$_2$S. The silica surface on the grid was functionalized with reactive amine groups which might react with the H$_2$S \cite{OkonkwoCEJ20} and carry the sulfide to the next step. The sample Ag NPs were in contact with the silica + amine surface in both polar and apolar solvent, and also under light irradiation during the measurement, which can be important for the reactivity of silver. Oxidation of silver can also be promoted by UV light.
Generally speaking, it is well known that tarnish commonly forms on silverware when left in atmosphere for extended periods of time. This by itself indicates that the atmospheric sulfur or oxygen content is enough to promote tarnishing of exposed silver surfaces, with no need of being fostered by reactions specific to our protocol. Even more so in the case of silver in NP form, whose reactivity is increased by the high surface-to-volume ratio \cite{McmahonAPB05,ElechiguerraCM05}. 

\clearpage

\section{Correction for finite region of detection\label{sec:finite_detection}}
In our micro-spectroscopy experiments the imaged area is delimited along the dispersive direction by the input slit of the spectrometer having a width of 80\,\textmu m. The slit is imaged with same size onto the sensor, where it is matched along the orthogonal direction by the on-chip binning (we read out a bin of 5 pixels of 16\,\textmu m pitch).
Considering the magnification of about 80$\times$ from sample to sensor (characterized experimentally by a controlled displacement of the sample stage), the $80 \times 80$\,\textmu m region of interest on the sensor corresponds to a square imaged area of lateral size $L=1.0$\,\textmu m on the sample.
This value was chosen to accommodate a particle image -- which for sufficiently small particles corresponds to the \ac{psf} of the imaging system -- while leaving some margin for possible lateral drifts over the acquisition time (few tens of seconds; the typical thermal drift of the imaged position is about 100\,nm/min).
\par
However, the mathematical \ac{psf} of a point source extends infinitely in space, albeit in practice only few rings (if any) are typically visible above the background noise level.
This means that only a fraction $f<1$ of the particle signal is detected as the tails of the \ac{psf} are cropped by the spatial filtering of the image.
Note that $\BF{f}\neq\DF{f}$, since the \ac{psf} in \ac{bf} and \ac{df} images differ due to the different angular range of excitation and the different contrast mechanism.
Specifically, in \ac{df}, the scattered intensity is measured, with a \ac{psf} determined by the objective \ac{na} and particle focus (within an approximated scalar diffraction theory neglecting the polarization dependence) whereas in \ac{bf}, the transmitted power is measured, which results from the interference between incident and scattered field, leading to a partially coherent imaging.
Matching condenser and objective \ac{na}, as we do in our experiment, the \ac{psf} in \ac{bf} is of similar size as the one in \ac{df}.
\par
In our quantitative analysis, the reduction of the excitation and scattering signal due to the finite area of detection is accounted for by rescaling \sig{ext}{} and \sig{sca}{} by \BF{f} and \DF{f}\!, respectively.
We determine \BF{f} and \DF{f} for our set-up with the following procedure.
Widefield images are acquired with a low-noise scientific sCMOS camera (PCO Edge 5.5).
Illumination is provided by a 100\,W halogen lamp (Nikon V2-A LL 100\,W), filtered using bandpass filters (Thorlabs FKB-VIS-40) with centre wavelengths of [450;500;550;600]\,nm, so to address the wavelength dependence of $f$. The illumination and detection \ac{na} ranges for \ac{bf} and \ac{df} are the same as in the experiment (namely, we use the same condenser and set of 3D printed apertures, and the same objective) to ensure we characterize the same \ac{psf}. We also use the same silver nanohedra sample, although ideally the \ac{psf} is the same for any isotropic subwavelength object.
\par
\begin{figure}[tbp]
	\includegraphics[width=0.496\textwidth]{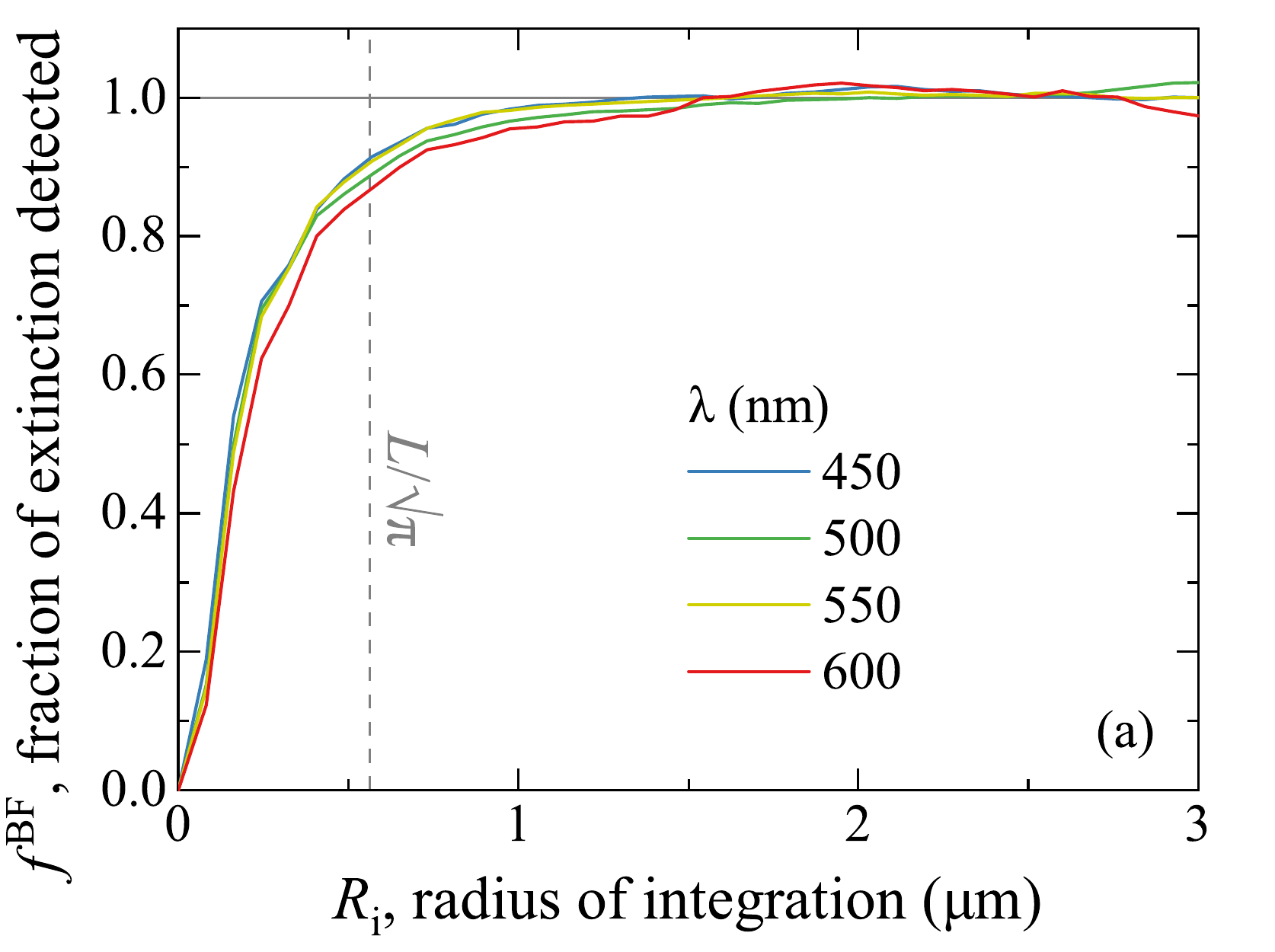}
	\hfill
	\includegraphics[width=0.496\textwidth]{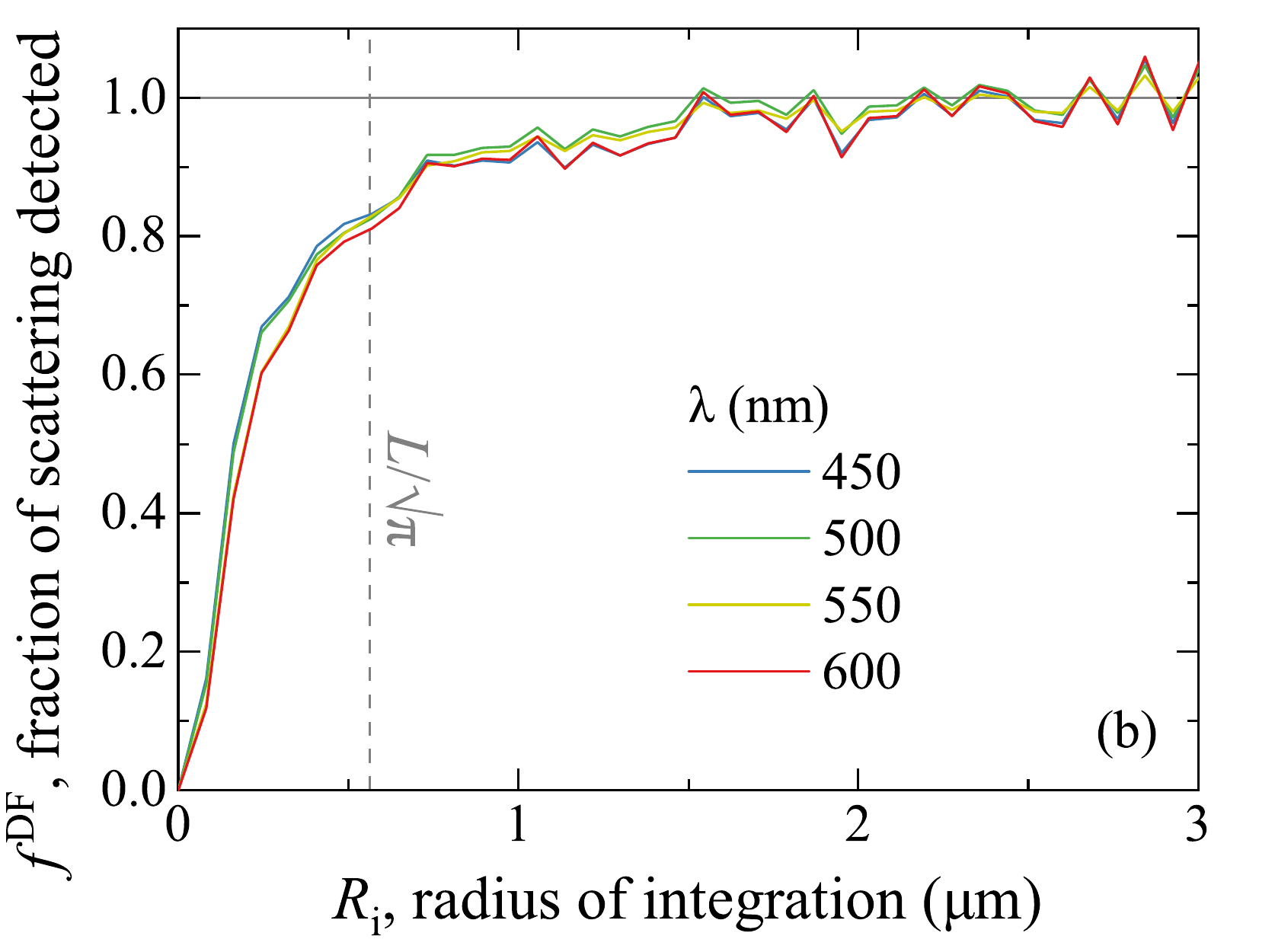}
	\caption{\label{fig:fBF_fDF}
		Fraction of (a) extinction and (b) scattering detected in imaging as a function of the integration radius \Ri, normalized to its saturation value at large \Ri{} indicated by the horizontal guideline at $f = 1$.
		The vertical dashed lines correspond to the equivalent radius of the imaged sample region in our microspectroscopy experiments.
	}
\end{figure}

We analysed the acquired transmission and scattering images using \emph{Extinction Suite}, a plug-in for the image processing programme ImageJ which we have been developing within our group -- see \url{https://langsrv.astro.cf.ac.uk/Crosssection/Crosssection.html} and publications referenced therein.
An analysis routine determines the particle position via a Gaussian fit of its transmission or scattering image; the extinction or scattering magnitude is quantified by integrating over a circular region of interest of radius \Ri{} centred around the particle position.
\autoref{fig:fBF_fDF}a shows a measurement of the extinction as a function of \Ri{}, after subtraction of the local background measured over an area $\pi\Ri^2$.
The extinction saturates at about $\Ri = 3\lambda/\mathrm{NA}\simeq 1.7$\,\textmu m, above which fluctuations of the local background dominate.
The scattering magnitude shown in \autoref{fig:fBF_fDF}b exhibits a similar behaviour with a slightly slower saturation (at about 2\,\textmu m).
We associate the value $f=1$ to the saturation magnitude -- indicated by the horizontal lines in \autoref{fig:fBF_fDF}a,b -- and normalize to it the extinction or scattering.
\begin{figure}[tbp]
	\includegraphics[width=.55\textwidth]{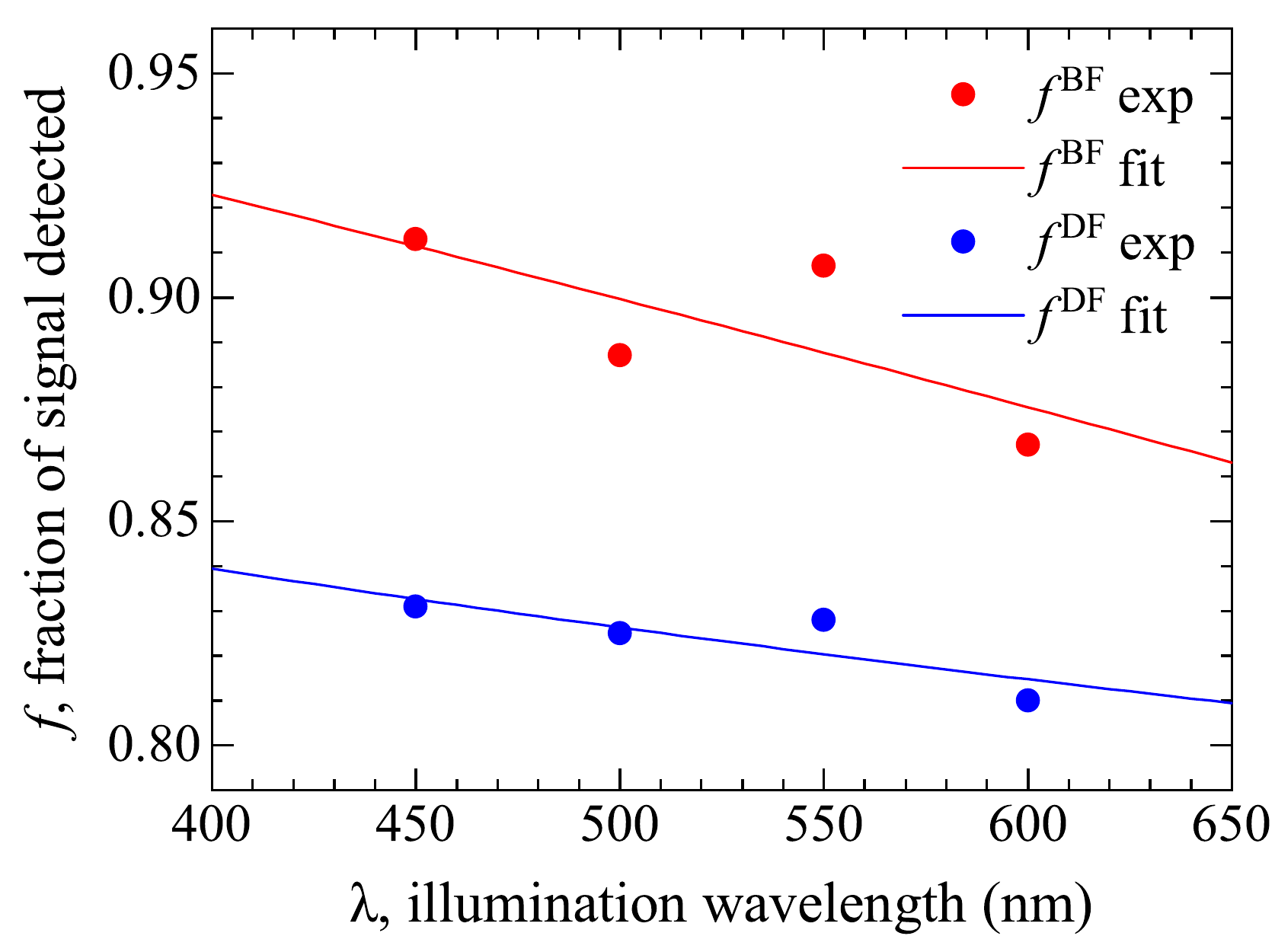}
	\caption{\label{fig:f_vs_lda}
		Measured values of \BF{f} and \DF{f} in four colour channels (circles).
		The wavelength dependence is fitted by a power law (lines) given by \autoref{eq:f_fit}.
	}
\end{figure}
\par
We estimate \BF{f} and \DF{f} in our micro-spectroscopy experiments at the equivalent radius $L/\sqrt{\pi} = \SI{564}{\nm}$ (vertical dashed line) which has the same area as the square region detected in micro-spectroscopy; the resulting values are reported in \autoref{fig:f_vs_lda} for the four colour channels used.
The experimental data are fitted with the phenomenological function
\begin{equation}
	\label{eq:f_fit}
	f(\lambda) = 1-(\lambda/A)^p
\end{equation}
and the parameters ($A = 4091$\,nm, $p = \num{1.18}$) for \ac{bf} and ($A = 71908$\,nm, $p = \num{0.353}$) for \ac{df}.
The functions $\BF{f}(\lambda)$ and $\DF{f}(\lambda)$ are used to correct the measured cross section magnitudes according to \autoref{eq:cross-sections} below.
The decreasing trend for longer $\lambda$ observed in in \autoref{fig:f_vs_lda} is explained by the scaling of the \ac{psf} with $\lambda$; for instance, the Airy function (which describes the focal spot created by a perfect lens with a circular aperture in the paraxial approximation) has a first dark ring of diameter $\num{1.22}\lambda/\mathrm{NA}$.
Note that this scaling is consistent with the limit behaviour $f\to 1$ for $\lambda\to 0$ of the fitting function of \autoref{eq:f_fit}. We estimate that the error in the determined factors is about 5 to 10\%, mostly due to the determination of the saturation value for large \Ri, which is affected by fluctuations in the background value that increase for larger integration areas.
\clearpage
\section{Characterization of the BF-to-DF illumination intensity\label{sec:xi}}
Another crucial parameter of the experimental set-up for quantifying the cross-section magnitude is the \ac{bf}-to-\ac{df} ratio of the illumination intensity.
The need for this parameter arises because in \ac{df} the excitation intensity cannot be directly measured, and it has therefore to be retrieved from the \ac{bf} background through a proportionality factor.
We call this parameter $\xi$, and it acts as a scaling factor for the magnitude of \sig{sca}{df} -- see \autoref{eq:cross-sections} below.
$\xi$ is governed by the amount of light blocked by the \ac{bf} and \ac{df} apertures in the \ac{bfp}.
It is therefore possible to derive a simple analytical expression of $\xi$ (Eq.~(3) in \Onlinecite{ZilliACSP19}) assuming an aplanatic behaviour of the condenser lens.
However, in most microscopy set-ups the illumination is not homogeneous over the \ac{bfp} of the condenser; moreover the condenser transmittance drops towards the edges of is aperture.
These effects add up to give a strong decrease of the illumination intensity at large \ac{na} values, which effectively lower the \ac{df} illumination.
Such angular efficiency of the excitation path can be characterized experimentally and used to correct $\xi$ as described in \S\,S.VI.B in \Onlinecite{WangNSA20}
\par
In this work (similar to what we did already in \Onlinecite{ZilliACSP19} for the polystyrene beads) we have instead measured $\xi=\num{2.04}$ directly for the specific \ac{bf} and \ac{df} 3D-printed apertures used in the experiment with the following procedure. Using an excitation path replicating the microspectroscopy experiments, a 1.45\,NA objective is used in the detection path to collect all exciting light also in the \ac{df} illumination configuration, which has a maximum of  1.34\,NA. A clean glass slide is used in place of the sample (to hold the immersion oil of objective and condenser) and K\"{o}hler illumination is adjusted focussing the field aperture. To minimize the effect of chromatic aberrations and reproduce the experimental focussing conditions a  colour filter centred at 550\,nm and a width of 40\,nm (Thorlabs FBH550-40) is used, which is the spectral region of the plasmonic resonance of the decahedra. Widefield images are then acquired with a scientific sCMOS camera (PCO Edge 5.5) using the \ac{bf} and \ac{df} 3D-printed apertures. The illumination intensity is proportional to the mean value of the camera readout over a region of interest in the centre of the field of view. Taking the ratio of \ac{bf} to \ac{df} readout (after subtracting from both the dark offset of the camera digitizer) yields $\xi$. Note that this procedure neglects the angular dependence of the collection efficiency; however, for the high-quality objective used (Nikon MRD01905, 100$\times$ 1.45\,NA PlanApo Lambda series) we expect this dependence to be weak over the range up to 1.34\,NA of the transmitted illumination, considering the significant margin to the objective NA.

\section{Spectroscopy of all particles\label{sec:all_spectra}}
Let us report here for convenience the formulas derived in \Onlinecite{ZilliACSP19} that we use to quantify the cross section magnitude, slightly adapted to match the notation of this work

\begin{subequations}
	\vspace{1em}
	\begin{minipage}{.4\textwidth}
		\begin{equation}
			\sig{sca}{df} = \frac{L^2}{\DF{f}} \frac{\xi}{\DF{\eta}} 
			\frac{\signal{\mathsc{np}}{df} - \signal{bg}{df}}{\signal{bg}{bf}}
			\tag{\theequation s}
			\label{eq:sig_sca}
		\end{equation}
	\end{minipage}%
	\hfill
	\begin{minipage}{.56\textwidth}
		\begin{equation}
			\sig{abs}{bf} = \frac{L^2}{\BF{f}} 
			\frac{\signal{bg}{bf} -\signal{\mathsc{np}}{bf}}{\signal{bg}{bf}}
			- (1 - \BF{\eta}) \frac{\zeta}{\xi} \sig{sca}{df}
			\tag{\theequation a}
			\label{eq:sig_abs}
		\end{equation}
	\end{minipage}
	\label{eq:cross-sections}
	\vspace{1em}
\end{subequations}

where $S(\lambda)$ are the the scattering/extinction spectra detected (after subtraction of the dark offset of the CCD digitizer) under \ac{df}/\ac{bf} illumination, imaging either the nanoparticle (NP subscript) or the background (bg subscript) in an empty area nearby.
The parameters $f$, $L$, and $\xi$ are specific to the experimental set-up and settings only (they have the same value for all measured particles) and were discussed in \autoref{sec:finite_detection} and \autoref{sec:xi}.
Conversely, the parameters $\zeta$ and $\eta$ are specific to each particle and, as mentioned in the article, encode the directional properties of scattering with respect to excitation and detection, respectively.
\par
\begin{figure}[p]
	\includegraphics[width=0.98\textwidth]{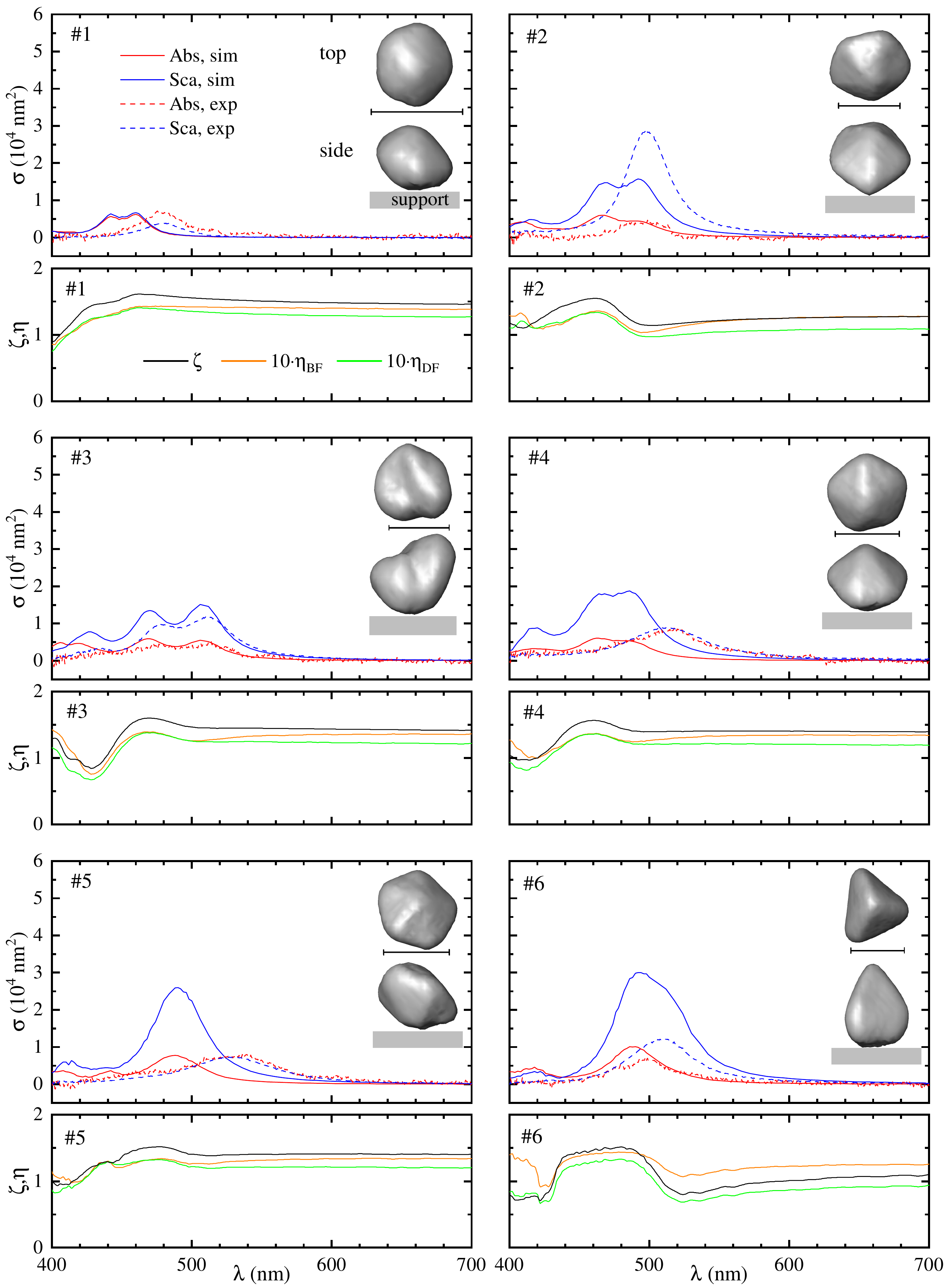}
	\caption{\label{fig:particles1to6}
		Measured and simulated optical cross section spectra (top) and parameters $\zeta$ and $\eta$ (bottom) of the particles \#1 to \#6.
		The insets are tomographic reconstructions (scale bar is 40\,nm).
	}
\end{figure}
\begin{figure}[p]
	\includegraphics[width=0.98\textwidth]{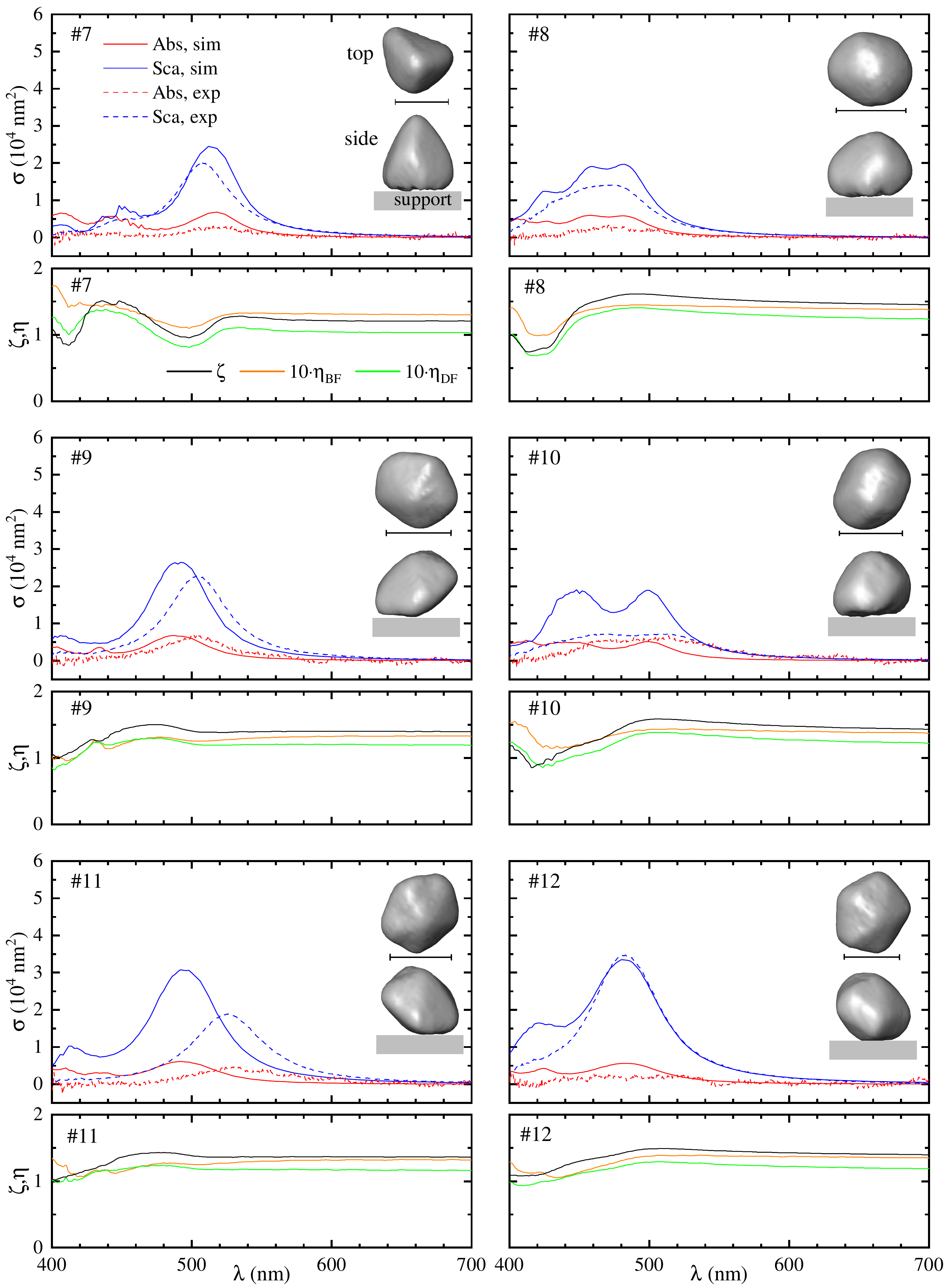}
	\caption{\label{fig:particles7to12}
		Same as \autoref{fig:particles1to6}, but for particles \#7 to \#12.}
\end{figure}
\begin{figure}[p]
	\includegraphics[width=0.98\textwidth]{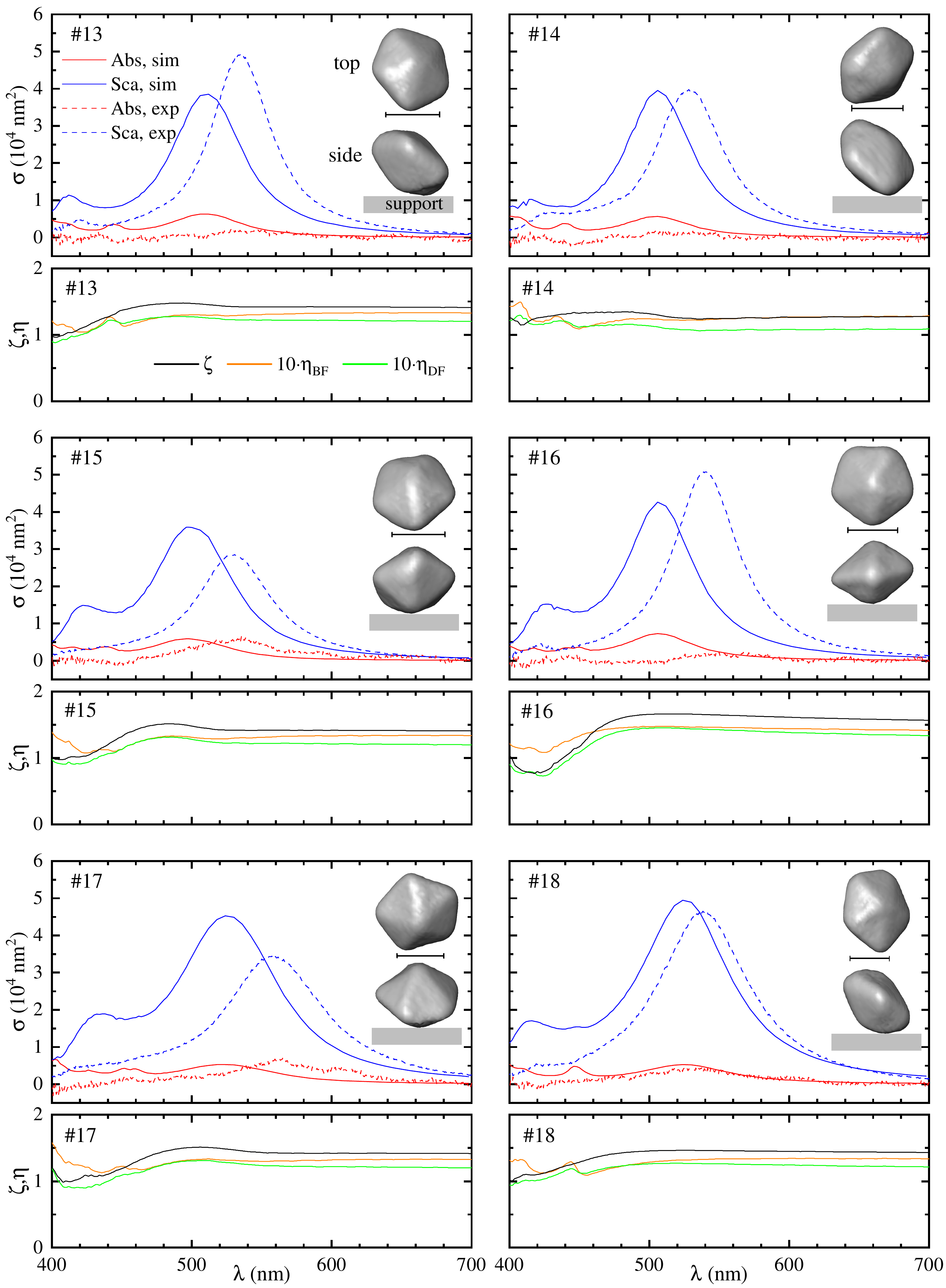}
	\caption{\label{fig:particles13to18}
		Same as \autoref{fig:particles1to6}, but for particles \#13 to \#18.}
\end{figure}
\begin{figure}[t]
	\includegraphics[width=0.98\textwidth]{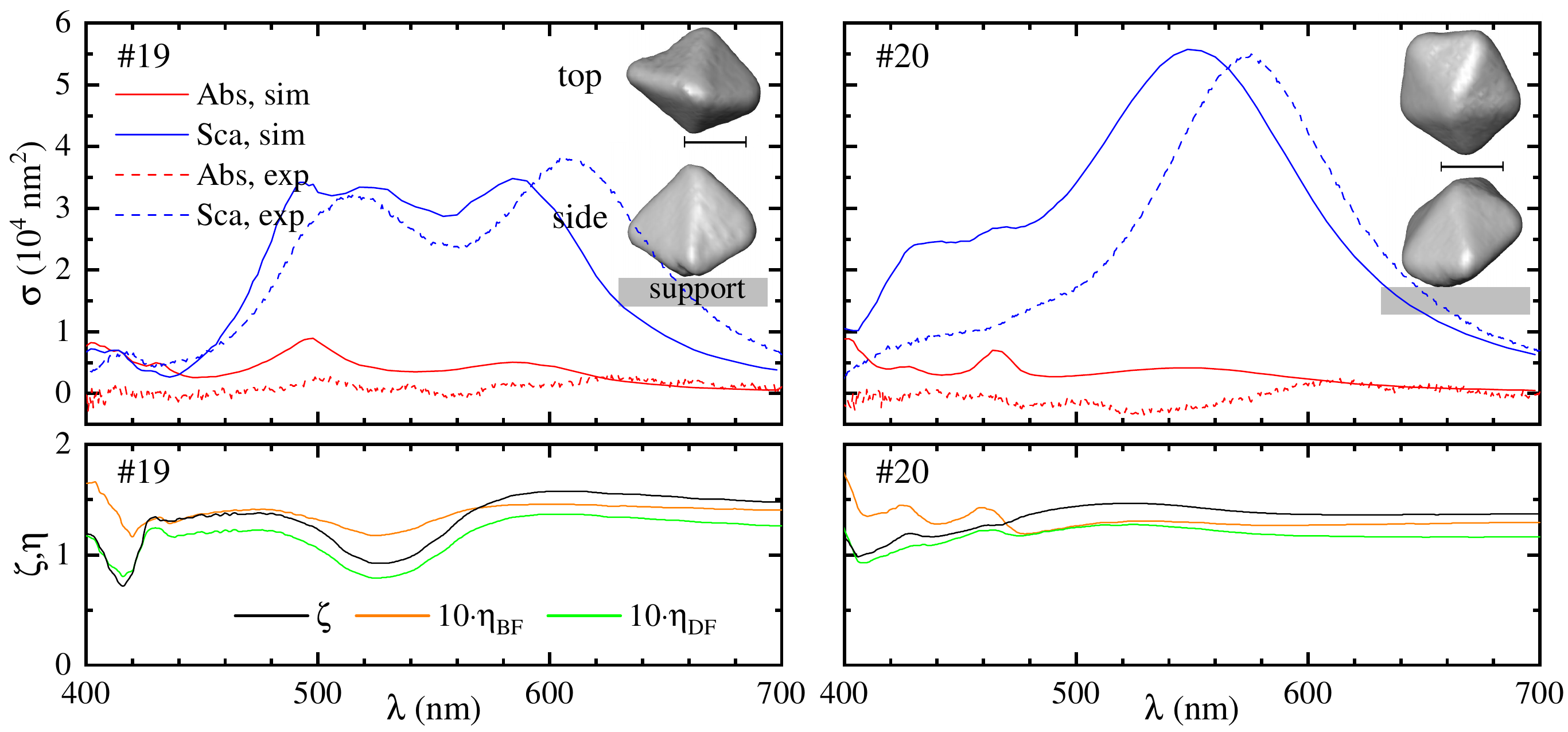}
	\caption{\label{fig:particles19to20}
		Same as \autoref{fig:particles1to6}, but for particles \#19 and \#20.}
\end{figure}

\autoref{P-fig:fig2} of the article shows the measured and simulated quantitative optical cross-section spectra $\sigma(\lambda)$ of six representative particles. The data for all twenty particles investigated in this work are shown in \autoref{fig:particles1to6}--\ref{fig:particles19to20}; a \ac{tem} tomographic reconstruction of each particle is included as an inset.
The study we presented in the article focuses on $\sig{sca}{df}(\lambda)$ which dominates the response of the investigated particles.
The untreated absorption spectra (not shown here) display for most particles negative values over a large spectral range in correspondence of the dipolar plasmonic resonance governing the scattering spectra.
To pinpoint the origin of such non-physical result (which would imply a net power emission from the particle) it is useful to look at the structure of \autoref{eq:sig_abs}.
The first term is the measured \sig{ext}{bf} and the second term is the portion of \ac{bf} scattering not collected by the objective, which is subtracted from the total extinction to isolate the absorptive contribution.
Therefore $\sig{abs}{bf} < 0$ can result from either underestimating \sig{ext}{bf} or overestimating $\sig{sca}{bf} = (\zeta/\xi)\sig{sca}{df}$.
The first can be corrected by decreasing \BF{f} and the latter by increasing \DF{f} or decreasing $\zeta$.
Each correction corresponds to different aspects of uncertainty in the experiment; specifically, $f$ refers to spatial filtering (\eg{} the particle drifts away from the centre of the imaged area during the acquisition, and hence $f$ is lowered) and $\zeta$ to directional filtering (\eg accuracy of the fabrication and positioning of the 3D printed apertures in the \ac{bfp} of the condenser).
For the spectra shown in \autoref{fig:particles1to6}--\ref{fig:particles19to20} we decreased the parameter \BF{f} by 20\% to avoid negative values of the spectral average of the experimental absorption.
Note that for these particles the opposite contributions of extinction and scattering have similar magnitude, so that they approximately balance each other; this implies that a similar result can be obtained by a 20\% correction of either \DF{f} or $\zeta$. An even more accurate measurement of the parameters for the experimental setup seems required to avoid such an adjustment. We emphasize that the same correction was used for all spectra shown.
\par
Let us now turn our attention to the scattering parameters, $\zeta(\lambda)$ and $\eta(\lambda)$, which are computed for each particle and correlated to $\sigma(\lambda)$ in \autoref{fig:particles1to6}--\ref{fig:particles19to20} (bottom panels).
It is instructive to compare these numerical simulations -- which take into account the complexity of the particle shape -- to our previous analytical calculations performed in the electrostatic approximation -- see section S.V. of \Onlinecite{ZilliACSP19}.
Let us start with $\eta$, which is the fraction of the scattering power collected by the objective.
Given the excitation and detection \ac{na} ranges of the experiment and assuming a homogeneous immersion medium, one finds $\BF{\eta}=\DF{\eta}=\num{0.148}$ for a polarisability perpendicular to the optical axis and $\BF{\eta}=\num{0.136}$, $\DF{\eta}=\num{0.111}$ for an isotropic polarisability.
The simulated $\eta$ compares well with this estimate, being closer to the spherical value for most particles.
Note that $\eta$ is determined by the orientations of the electric dipoles excited in the particle: the larger the angle formed with the optical axis of the objective, the lower the fraction of emission collected, resulting in the largest $\eta$ for a dipole lying flat on the substrate.
Looking at the simulations below, these considerations also explain why $\BF{\eta}>\DF{\eta}$ -- as more inclined dipoles are excited in \ac{df} -- and why for most particle $\eta$ decreases for $\lambda < 450$\,nm, where multipolar resonances are excited.
\par
As for $\zeta$, which is the \ac{bf}-to-\ac{df} ratio of the total scattered power, no straightforward comparison to the values in the dipole limit ($\zeta=\num{1.22}$ for an in-plane polarisability and $\zeta=\num{0.897}$ for an isotropic polarisability) can be made.
This is because in this set of measurements the \ac{na} range of the \ac{df} illumination reaches the edge of the condenser aperture (1.34\,NA) where the illumination intensity is significantly reduced -- see Fig.~S7 of \Onlinecite{WangNSA20}.
The numerical modelling used in this work takes into account the angular dependence of the illumination intensity based on our experimental characterization of the performance of the condenser -- see \Onlinecite{WangNSA20}. Conversely, the analytical calculations assume a homogeneous filling of the back aperture of the condenser, thereby overestimating the \ac{df} illumination and scattered power, which leads to a lower $\zeta$.
Along the same line of reasoning, the dip of $\zeta$ for $\lambda<450$\,nm implies that the multipolar modes in that region are comparatively better excited by the tilted illumination of \ac{df}.
\clearpage
\section{Morphology reconstruction from HAADF-STEM tomography and its influence on the simulated cross-sections\label{sec:geometry_modification}}
\subsection{Remeshing\label{sec:remeshing}}
As described in the main text, we investigated the dependence of our results on the geometry reconstruction method, and evaluated three procedures (R1 to R3). The different reconstruction methods were applied to a selection of particles, and the resulting volumes, and volume-to-surface ratios are given in \autoref{tab:volumes} along with a summary of the parameters used for each algorithm. 
Generally we find that the volume vary by some 5\% to 10\%, with R1 resulting in the lower and R2 in the higher volumes. The volume to surface ratios also vary by some 5\% to 10\%, but there is no clear trend across the particles for the different reconstructions.

\begin{figure}[bt]
	\includegraphics[width=0.98\textwidth]{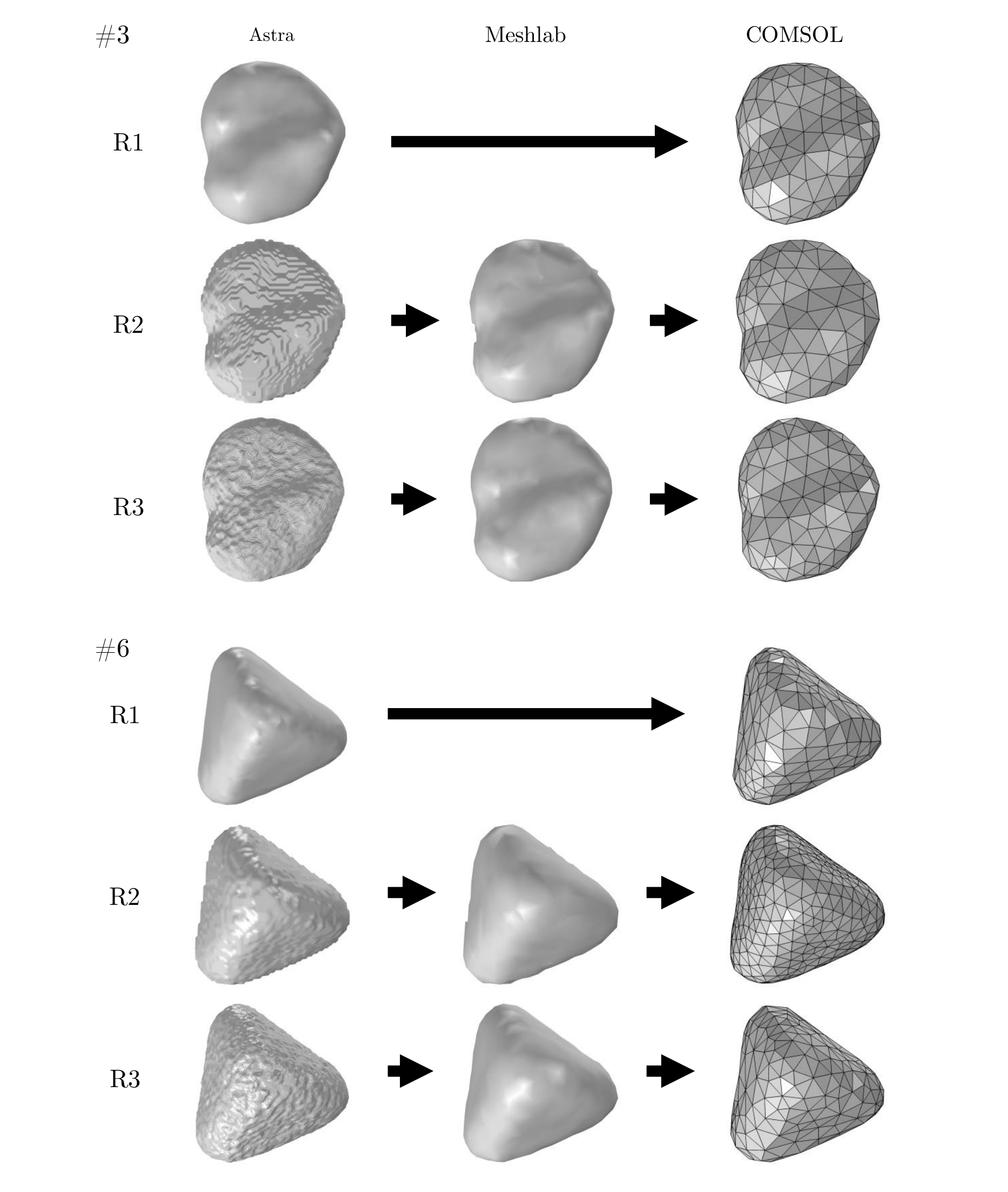}
	\caption{\label{fig:remesh_steps}
		Comparison of three geometry reconstruction procedures (R1 to R3) for two different particles (\#3 and \#6) viewed from the top.
		For R1 the 3D reconstruction is directly imported into \comsol and then meshed, while for R2 and R3 an intermediate step is introduced to reduce the number of faces defining the geometry.
	}
\end{figure}

While with the R1 procedure the resulting mesh can be directly imported and re-meshed by \comsol, R2 and R3 sizeably increase the number of surface elements defining the particle, which could not be imported, processed, and meshed reliably with \comsol. We therefore reduced the number of surface elements using the free software Meshlab and a procedure illustrated in \autoref{fig:remesh_steps} for two exemplary particles of different appearance. First the number of faces was changed to 1000 using the option `quadratic edge collapse', then the result was turned into pure triangular mesh, and finally the errors in the geometry (such as holes or crossing mesh elements) were repaired by the option `remove non manifold edges by removing faces'. The resulting mesh was then imported into \comsol{}. 
\begin{table}[bp]
	\centering
	\caption{\label{tab:volumes}
		Volume and volume-to-surface ratio of particles reconstructed with different procedures.
		The parameters identifying the reconstruction procedures R1, R2, R3 are given with the following abbreviations: It: iterations, N: factor of downsampling, Sm: smoothing, Rm: remeshing.}
	\vspace{0.8em}
	\begin{tabular}{|c|c|c|c|c|c|c|} 
		\hline
		\multirow{3}{*}{NP~~\#} & \multicolumn{2}{c|}{R1}                                                                           & \multicolumn{2}{c|}{R2}                                                                            & \multicolumn{2}{c|}{R3}                                                                              \\ 
		\cline{2-7}
		& \multicolumn{2}{c|}{\begin{tabular}[c]{@{}c@{}}It = 15, N = 12, \\Sm = No, Rm = No\end{tabular}} & \multicolumn{2}{c|}{\begin{tabular}[c]{@{}c@{}}It = 15, N = 4, \\Sm =Yes, Rm = Yes\end{tabular}} & \multicolumn{2}{c|}{\begin{tabular}[c]{@{}c@{}}It = 100, N = 1, \\Sm = Yes, Rm = Yes\end{tabular}}  \\ 
		\cline{2-7}
		& V (10\textsuperscript{4} nm\textsuperscript{3}) & V/S (nm)                                        & V (10\textsuperscript{4}~nm\textsuperscript{3}) & V/S (nm)                                         & V (10\textsuperscript{4}~nm\textsuperscript{3}) & V/S (nm)                                           \\ 
		\hline
		3                       & 4.09                                            & 6.65                                            & 4.29                                            & 6.79                                             & 4.10                                            & 6.50                                               \\ 
		\hline
		6                       & 4.67                                            & 5.88                                            & 5.24                                            & 5.19                                             & 5.20                                            & 7.07                                               \\ 
		\hline
		7                       & 4.92                                            & 6.87                                            & 5.42                                            & 7.24                                             & 5.33                                            & 7.12                                               \\ 
		\hline
		18                      & 12.2                                            & 9.53                                            & 12.3                                            & 9.53                                             & 11.9                                            & 9.30                                               \\ 
		\hline
		19                      & 12.6                                            & 9.13                                            & 14.6                                            & 9.80                                             & 14.0                                            & 9.46                                               \\ 
		\hline
		20                      & 18.4                                            & 10.8                                            & 18.2                                            & 10.6                                             & 17.4                                            & 10.4                                               \\
		\hline
	\end{tabular}
\end{table}

\clearpage
\subsection{Cross-section spectra}
\begin{figure}[b]
	\includegraphics[width=0.98\textwidth]{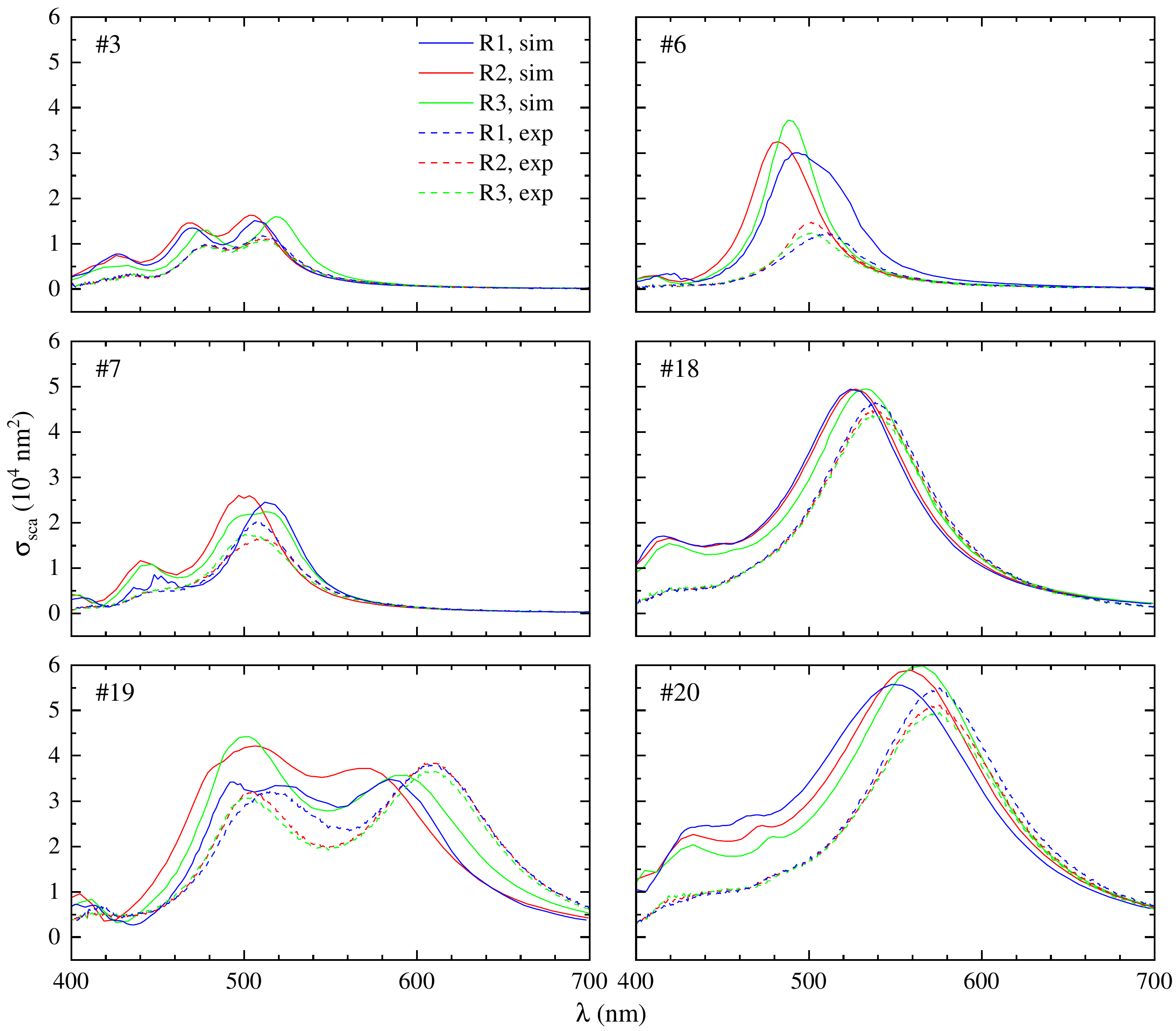}
	\caption{\label{fig:geometry}
		Simulated (solid lines) and experimental (dashed lines) scattering cross section spectra of the particles \#3, \#6, \#7, \#10, \#18, \#19, and \#20, as labelled, for different geometry reconstruction procedures R1 (blue lines), R2 (red lines), and R3 (green lines).}
\end{figure}

\autoref{fig:geometry} shows the scattering cross-section spectra of the six particles studied in the article obtained using the different \ac{tem} tomography reconstruction procedures R1 to R3. 
There is no clear common trend of the effect of the different reconstructions. We typically see variations of some 10 to 20\,nm in peak position, and in peak splitting, and some 5\% to 20\% in peak amplitude.  

\begin{figure}[h]
	\includegraphics[width=0.98\textwidth]{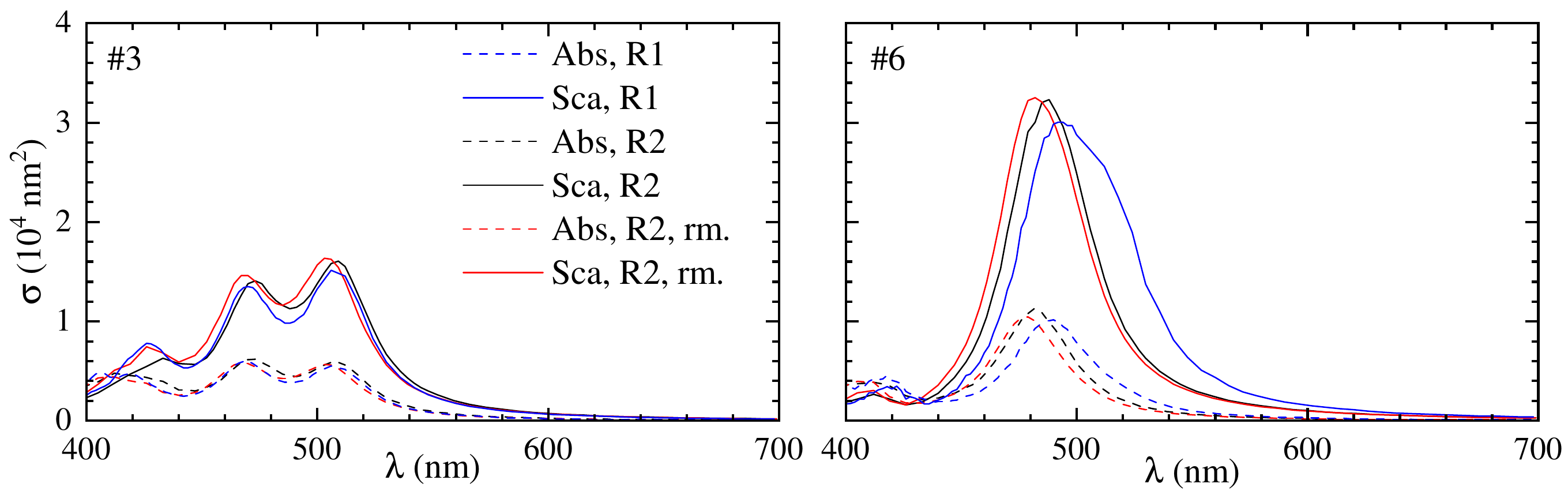}
	\caption{\label{fig:remeshing}
		Simulated cross-section spectra of particles constructed via the R2 algorithm, with (labeled rm.) and without remeshing before importing into \comsol.}
\end{figure}

\clearpage
For particle \#3 and \#6 \comsol could process the R2 geometries without remeshing, therefore we can use these two particles to investigate the effects of the remeshing. In \autoref{fig:remeshing} we show the simulated cross section spectra for these two particles. We find a small blue-shift of approximately 5\,nm due to the remeshing for both particles, and an increase in amplitude below 1\%. The effects should be even smaller for larger particles, which are less sensitive for small surface changes that are caused by the meshing.
All other simulations shown in the supplement or in the article uses the outlined remeshing steps for R2 and R3, and the `rm' label is dropped.

\subsection{Rotating three-dimensional renderings of the particle geometry}\label{subsec:3Dmovies}

We provide as online material animations of the reconstructed geometries of all reported NPs.
\clearpage
\section{Surface and interface Drude damping \label{sec:permittivity_modification}}
\begin{figure}[bp]
	\includegraphics[width=0.98\textwidth]{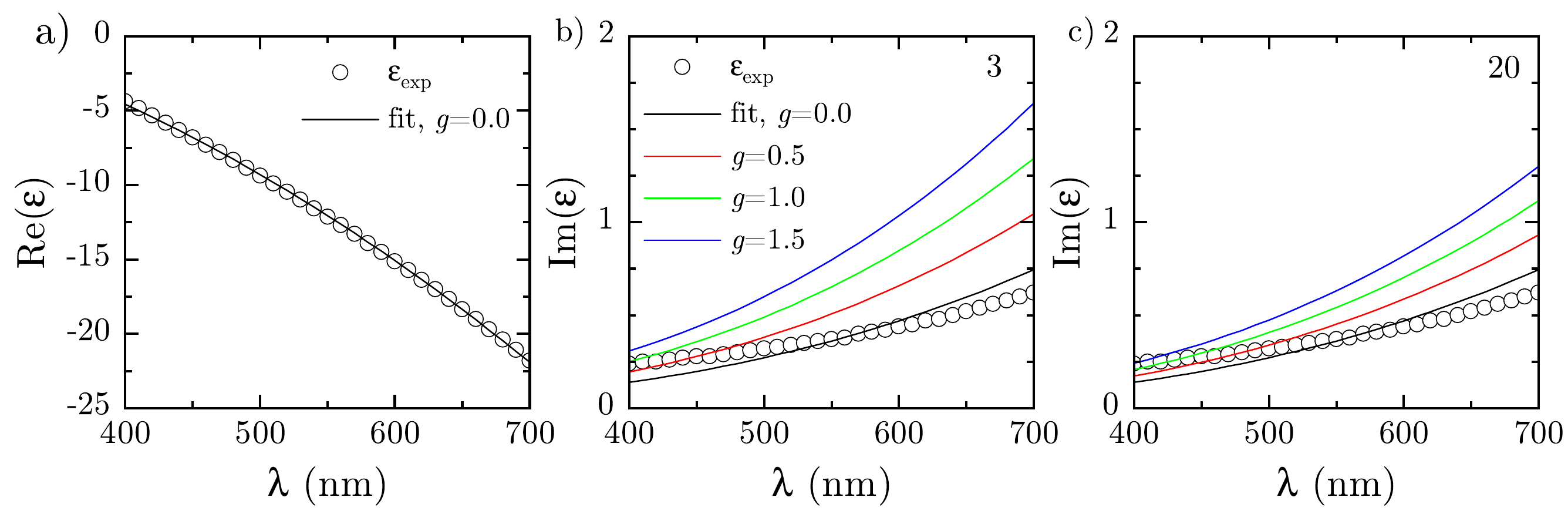}
	\caption{\label{fig:permittivity}
		Fit of the experimental permittivity dataset of \onlinecite{YangPRB15} with the Drude model and additional damping.
		a) Real part, data (circles), and model (line).
		b) Imaginary part, data (circles), and model (lines) for $g=0$ (black), as well as with the added damping using $g=\num{0.5}$ (red), 1.0 (green), and 1.5 (blue), for particle \#3.
		c) Same as b) but for particle \#20.
	}
\end{figure}
In this section we investigate the effect of increasing the Drude damping in the Ag permittivity, to model the increased surface or defect scattering in the particles compared to the permittivity datasets \cite{YangPRB15} measured by ellipsometry on thin films.
Such an increase is expected due to the particle size being smaller than the crystallite sizes in the measured films, and the additional crystal defects which can be created in the colloidal growth \cite{ZhengL09,XuCM21}.
We fit the data set \cite{YangPRB15} with the Drude model $\epsilon(\omega, \gamma) = \epsilon_\infty - \omega_\mathrm{p}^2 / (\omega^2 + i\omega \gamma)$ in the range 400 to 700\,nm.
The fit parameters are $\epsilon_\infty = 3.8575$, $\omega_\mathrm{p} = 1.3666e16$\,s$^{-1}$ and $\gamma= 7.7849e13$\,s$^{-1}$.
The resulting analytical permittivity is shown in \autoref{fig:permittivity} along with the fitted experimental dataset $\epsilon_\mathrm{exp}(\omega)$ of \Onlinecite{YangPRB15}. We note from panel b that the imaginary part of the permittivity has some deviations from the Drude model in this range, and it has been shown \cite{SehmiPRB17} that additional poles are needed for an accurate fit. However, since we are here only interested in modelling the change of the permittivity by increasing Drude damping, the simpler model suffices.  
Following \cite{VoisinJPCB01}, we then add a damping $\Delta \gamma= g \vF/R$ where $\vF = 1.36e6$\,m/s is the Fermi velocity, and the equivalent radius $R$ is calculated from the volume based on a spherical particle, $R= \sqrt[3]{3V/4\pi}$, and replace $\gamma$ with $\ovl{\gamma} = \gamma + \Delta \gamma$ in the modified permittivity $\epsilon(\omega, \ovl{\gamma}) = \epsilon_\infty - \omega_\mathrm{p}^2 / (\omega^2 + i \omega \ovl{\gamma})$.
We vary the damping parameter $g$, and the resulting change to the imaginary part of the permittivity $\epsilon(\omega, \ovl{\gamma})$ is shown in \autoref{fig:permittivity}b,c.
For particle \#20 the permittivity changes less compared to \#3 due to its larger $R$.
The real part of the permittivity is changed by less than 0.1\% therefore this is neglected here.
We take the change of the Drude permittivity $\Delta\epsilon(\omega, \ovl{\gamma}) = \epsilon(\omega, \ovl{\gamma}) - \epsilon(\omega, \gamma)$, and add it to the measured data $\epsilon_\mathrm{exp}(\omega)$, resulting in the modified permittivity  $\epsilon_{\rm m}(\omega, \ovl{\gamma}) = \epsilon_\mathrm{exp}(\omega) + \Delta\epsilon(\omega, \ovl{\gamma})$ used in the simulation.
\par
\begin{figure}[tbp]
	\includegraphics[width=\textwidth]{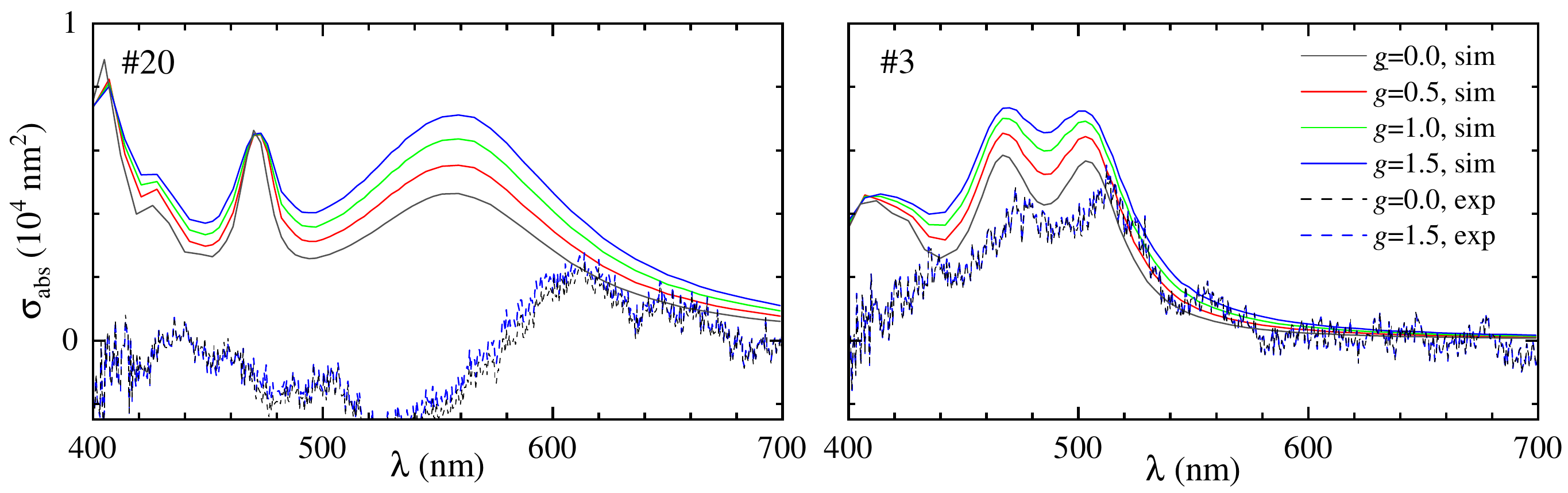}
	\caption{\label{fig:absorption}
		Simulated (solid lines) and measured (dashed lines) absorption cross section spectrum of particle \#20 (left) and \#3 (right) for different magnitudes $g$ of the surface damping.
	}
\end{figure}

In the article we show in \autoref{P-fig:fig4} the effect of the permittivity change on the scattering cross section, while \autoref{fig:absorption} shows the effect on the absorption cross section.
The simulated absorption increases for stronger damping as expected.
The experimental absorption, like already seen for the scattering, is unaffected because the simulated parameters $\eta$ and $\zeta$ are only weakly affected by the increased damping (in the electrostatic limit, they are dispersionless and depend only on the particle geometry but not its material properties.).

\section{Cross-section spectra for sulfide or oxide tarnishing \label{sec:tarnish_composition}}
\begin{figure}[t]
	\includegraphics[width=0.65\textwidth]{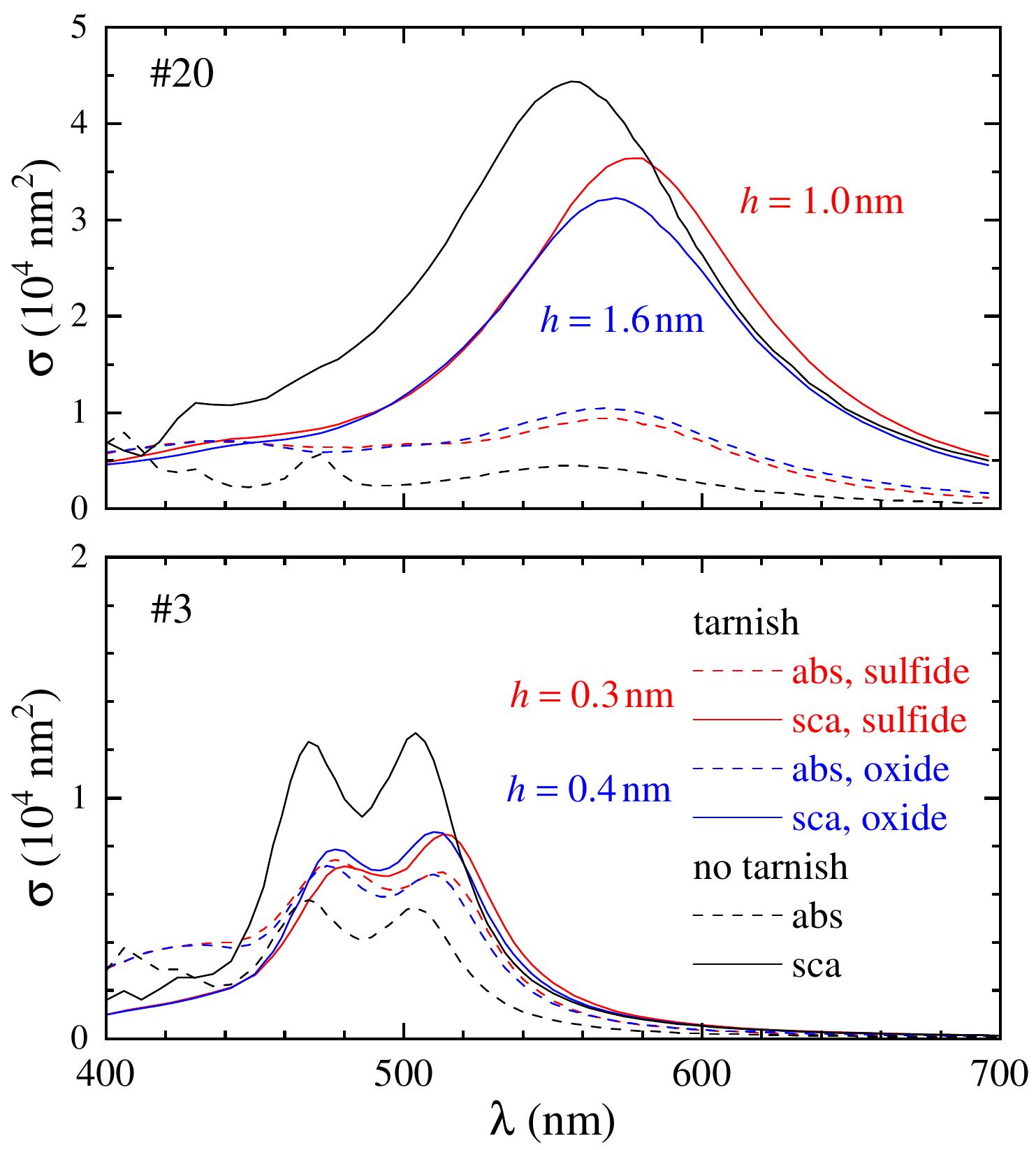}
	\caption{\label{fig:tarnish_compare}
		Simulated cross-section spectra of particle \#20 (top) and \#3 (bottom) for a silver-sulfide (red) and a silver-oxide (blue) tarnish layer of thickness $h$ as given covering the particle, for normal incidence illumination. 
	}
\end{figure}

We have already discussed in \autoref{sec:sulfidisation} that the chemical composition of the tarnish layer on the NP surface is uncertain, although silver sulfide (Ag$_2$S) seems the most likely candidate based on previous reports in literature. In this section, we investigate a possible different composition of such layer, namely silver oxide (Ag$_2$O), comparing the simulated spectra to those obtained with an Ag$_2$S layer, which was used in the main text for the NPs \#3 and \#20 -- see \autoref{P-fig:fig6}.

The permittivity spectra used as material properties were taken from Ref.\,\cite{BennettJOSA70} for Ag$_2$S as in the main text, and from Ref.\,\cite{PetterssonTSF95} for Ag$_2$O. To compare the effect of the two materials, it is sufficient to evaluate the cross-section spectra for normal incidence illumination, given that the cross-section spectra quantitatively simulating the measurements, which use a range of illumination directions, are reported for  Ag$_2$S tarnish layers in the main text. The layers are modelled as described in \autoref{P-ss:tarnish}. 

The resulting cross-section spectra are shown in \autoref{fig:tarnish_compare}. For the Ag$_2$S tarnish layer, the same thicknesses as in the main text are used, while for the Ag$_2$O tarnish layer, the thicknesses are chosen to provide a similar change of the cross-sections as for the Ag$_2$S tarnish layer.
For particle \#20, the scaling factor for the Ag$_2$S layer is 0.97, yielding a thickness of approximately 1\,nm. The layer redshifts the dipolar peak and decreases its amplitude, as discussed in the main text. For the Ag$_2$O layer, a scaling factor of 0.95 was used, yielding a thickness of about 1.6\,nm. We find that the Ag$_2$O layer results in a slightly larger amplitude reduction for a given shift. This would slightly increase the deviation from the experiment seen in \autoref{P-fig:fig6} for the Ag$_2$S layer. For particle \#3 , the scaling factor for the Ag$_2$S layer is 0.985,  yielding a thickness of approximately 0.3\,nm. For the Ag$_2$O layer, a scaling factor of 0.98 was used, yielding a thickness of about 0.4\,nm. Here the lower shift for the oxide layer would slightly decrease the deviation from the experiment. The actual morphology of the tarnish is likely more complex than the thin layer of homogeneous thickness used here for modelling; for example one could expect a higher reactivity of corners. Therefore, the observed small differences between Ag$_2$S and Ag$_2$O layers are not conclusive. Notably, Ag$_2$O could not be detected in the EDX results (see \autoref{sec:sulfidisation}) due to the presence of oxygen in the SiO$_2$ support, so that even the Ag$_2$O thickness of 1.6\,nm used for particle \#20 would not be easily visible in EDX.  

\clearpage
%
%